\documentclass[final,1p,times,english]{elsarticle}

\usepackage{graphicx}
\usepackage{subfigure}
\usepackage{tikz}
\usepackage{pgfplots}
\usepackage{hyperref}
\hypersetup{
  colorlinks,
  citecolor=Violet,
  linkcolor=Red,
  urlcolor=Blue}
\usepackage{amsthm}

\usepackage{multirow}
\usepackage{color}

\usepackage[bold]{hhtensor}

\biboptions{authoryear,semicolon,round}

\usepackage[T1]{fontenc}
\usepackage[latin9]{inputenc}
\usepackage{units}
\usepackage{babel}

\journal{International Journal of Plasticity}

\begin{document}
\newcommand{\etal}{{\it et al.}}

\begin{frontmatter}



\title{Thermally-activated Non-Schmid Glide of Screw Dislocations in W using Atomistically-informed Kinetic Monte Carlo Simulations}


\author[add1,add2]{Alexander Stukowski}
\author[add2,add5]{David Cereceda}
\author[add3,add4]{Thomas D. Swinburne}
\author[add2]{and Jaime Marian}
\address[add1]{Institute of Materials Science, Darmstadt University of Technology, Darmstadt D-64287, Germany}
\address[add2]{Physical and Life Sciences Directorate, Lawrence Livermore National Laboratory, Livermore, CA, USA}
\address[add5]{Instituto de Fusi\'on Nuclear, Universidad Polit\'ecnica de Madrid, E-28006 Madrid, Spain}
\address[add3]{Department of Physics, Imperial College London, Exhibition Road, London SW7 2AZ, United Kingdom}
\address[add4]{CCFE, Culham Science Centre, Abingdon, Oxon, OX14 3DB, UK}

\begin{abstract}
Thermally-activated $\small{\nicefrac{1}{2}}\langle111\rangle$ screw dislocation motion is the controlling plastic mechanism at low temperatures in body-centered cubic (bcc) crystals. Motion proceeds by nucleation and propagation of atomic-sized kink pairs susceptible of being studied using molecular dynamics (MD). However, MD's natural inability to properly sample thermally-activated processes as well as to capture $\{110\}$ screw dislocation glide calls for the development of other methods capable of overcoming these limitations. Here we develop a kinetic Monte Carlo (kMC) approach to study single screw dislocation dynamics from room temperature to $0.5T_m$ and at stresses $0<\sigma<0.9\sigma_P$, where $T_m$ and $\sigma_P$ are the melting point and the Peierls stress. The method is entirely parameterized with atomistic simulations using an embedded atom potential for tungsten. To increase the physical fidelity of our simulations, we calculate the deviations from Schmid's law prescribed by the interatomic potential used and we study single dislocation kinetics using both projections. We calculate dislocation velocities as a function of stress, temperature, and dislocation line length. We find that considering non-Schmid effects has a strong influence on both the magnitude of the velocities and the trajectories followed by the dislocation. We finish by condensing all the calculated data into effective stress and temperature dependent mobilities to be used in more homogenized numerical methods.
\end{abstract}


\end{frontmatter}


\section{Introduction}
\label{intro}

$\small{\nicefrac{1}{2}}\langle111\rangle$ screw dislocations are the main carriers of plasticity in body-centered cubic (bcc) single crystals. Experimentally, bcc slip is seen to occur on $\{110\}$, $\{112\}$, and $\{123\}$ planes, or any combination thereof.  To determine the slip plane for a general stress state, Schmid's law is used, which states that glide on a given slip system commences when the resolved shear stress on that system, the Schmid stress, reaches a critical value \citep{schmid1935}. This law is known to break down in bcc metals, with great implications for the overall plastic flow and deformation behavior in these systems. Experimentally, \emph{non-Schmid} behavior is well documented in the literature going back  several decades \citep{sestak1965,sherwood1967,zwi1979,christian1983,pichl2002}\footnote{Although it was first recognized as early as in the 1920s and 30s}, and its reasons have  been thoroughly investigated. First, as Vitek and co-workers have noted \citep{duesbery1998,ito2001}, slip planes in bcc crystals do not display mirror symmetry (a common characteristic of planes belonging to the $\langle111\rangle$ zone), and so the sign of the applied stress does matter to determine the critical stress. This is most often referred to as the \emph{twinning-antitwinning asymmetry}. Second, studies using accurate atomistic methods (semi empirical interatomic potentials and density functional theory calculations) have shown that stress components that are not collinear with the Burgers vector $\vec{b}$ couple with the core structure of screw dislocations resulting also in anomalous slip \citep{bulatov1999,rao2001,groger2005,chaussidon2006}.

Although, effective corrections that reflect deviations from Schmid law have been implemented in crystal plasticity models, and their effects assessed at the level of grain deformation \citep{dao1996,vitek2004,groger2005,yal2008,wang2011,lim2013,chen2013,soare2014}, there is no model establishing the fundamental impact of non-Schmid behavior on single screw dislocation motion. Molecular dynamics (MD) simulations naturally include non-Schmid effects as part of the simulated dynamics of screw dislocations \citep{gilbert2011,cere2013}. However, it is exceedingly difficult to separate these effects from the bundle of processes (and artifacts) brought about by size and time limitations inherent to MD simulations. In addition, screw dislocation motion proceeds by way of the nucleation and sideward relaxation of so-called \emph{kink pairs} in a broad stress and temperature range. Kink pair nucleation may be regarded as a rare event occurring on a periodic energy substrate known as the \emph{Peierls potential}. MD's inability to sample these events accurately often leads to overdriven  dynamics and unrealistically high dislocation velocities \citep{cere2013}.

Here, we develop a kinetic Monte Carlo (kMC) model to study thermally activated screw dislocation motion in tungsten (W). Our approach --which builds on previous works on the same topic \citep{lin1999,cai2001,cai2002,deo2002,PhysRevB.69.075209,pilar2012}-- is based on the stochastic sampling of kink pair nucleation coupled with kink motion. The entire model is parameterized using dedicated atomistic simulations using a state-of-the-art interatomic potential for W \citep{cosmin2013}. Non-Schmid effects are incorporated via a dimensionless representation of the resolved shear stress, which provides the deviation from standard behavior for all the different activated slip planes. We explore the impact of these deviations on single dislocation glide and compare the results to direct MD simulations. Another novel aspect of our simulations is the inclusion of stress-assisted kink drift and kink diffusion simultaneously in our model. This quantitatively reflects the behavior observed atomistically at the level of single screw dislocation motion.

The paper is organized as follows. First we describe the kMC algorithm and the topological construct of screw dislocations and kink segments. We then provide a detailed account of the parameterization effort undertaken, beginning with single kink static and dynamic properties, and ending with the calculation of the non-Schmid law. In the Results section we report calculations of Schmid and non-Schmid glide as a function of stress, temperature, dislocation length, and maximum resolved shear stress (MRSS) plane. We finish with a discussion of the results and the conclusions. 

\section{Kinetic Monte Carlo Model of Thermally-activated Screw Dislocation Motion}
\label{comp}

\subsection{Physical Basis}

All that is required to initialize a kMC run are the total initial screw dislocation line length $L$, the temperature $T$, and the applied stress tensor $\matr{\sigma}$. In the kMC calculations, we choose a working representation of the stress tensor in its non-dimensional scalar form:
$$s=\frac{\sigma_{\rm RSS}}{\sigma_P}$$
where $\sigma_{\rm RSS}$ is the \emph{resolved shear stress} (RSS) and $\sigma_P$ is the Peierls stress. We consider two different contributions to $\sigma_{\rm RSS}$, namely, from external sources --defined by an applied stress tensor $\matr{\sigma}$-- and from internal stresses originating from segment-segment elastic interactions. At a given dislocation segment $i$, the normalized resolved shear stress is:
\begin{equation}
s_i=\frac{\sigma_{ext}+\sigma_{int}}{\sigma_P}=\frac{\vec{t}\cdot\matr{\sigma}\cdot\vec{n}+\sum_{j}{\sigma_{ij}(\vec{r}_{j}-\vec{r}_i)}}{\sigma_P}
\label{s0}
\end{equation}
Here, $\vec{t}$ and $\vec{n}$ are unit vectors representing the slip direction and the glide plane normal, and $\vec{r}_i$ is the position of dislocation segment $i$. The calculation of $\sigma_{ij}$ is discussed in Section \ref{impl} but note that this definition of $\sigma_{int}$ introduces a certain locality in $s_i$, hence the subindex $i$. 

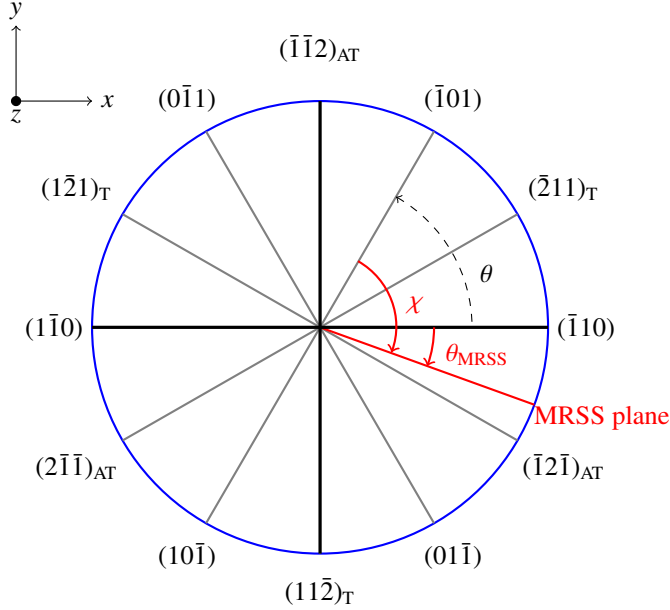
\begin{figure}[h]
\centering
  \begin{tikzpicture}
  \draw[blue,thick] (0,0) circle (3);    
    \foreach \a in {0, 30,...,359}
      \draw[thick,gray] (0, 0) -- (\a:3);
      
     \draw[thick,red] (0,0) --(-20:3); 
      \draw[red] (-18:3.9) node {MRSS plane};
      \draw[red] (13:1.3) node {$\chi$};
      \draw[red] (-10:2.1) node {$\theta_{\rm MRSS}$};
      \draw (18:2.3) node {$\theta$};
    \foreach \a in {0, 90,...,359}
      \draw[very thick] (0, 0) -- (\a:3);
  \draw (0: 3.5) node {$(\bar{1}10)$};
  \draw (180: 3.5) node {$(1\bar{1}0)$};
    \draw (270: 3.5) node {$(11\bar{2})_{\rm {\small T}}$};
  \draw (90: 3.7) node {$(\bar{1}\bar{1}2)_{\rm {\small AT}}$};
    \draw (60: 3.5) node {$(\bar{1}01)$};
  \draw (30: 3.7) node {$(\bar{2}11)_{\rm T}$};
      \draw (300: 3.5) node {$(01\bar{1})$};
  \draw (330: 3.7) node {$(\bar{1}2\bar{1})_{\rm AT}$};
        \draw (120: 3.5) node {$(0\bar{1}1)$};
\draw (150: 3.7) node {$(1\bar{2}1)_{\rm T}$};
    \draw (210: 3.7) node {$(2\bar{1}\bar{1})_{\rm AT}$};
    \draw (240: 3.5) node {$(10\bar{1})$};
\draw[->] (-4,3) -- (-3,3);
\draw[->] (-4,3) -- (-4,4);
\draw (-2.8,3) node {$x$};
\draw (-4,4.2) node {$y$};
\draw (-4,2.8) node {$z$};
 \draw[fill=black] (-4,3) circle(0.6mm);
   \draw[thick,red,->] (1.5,.0) arc (0:-20:1.5);
   \draw[thick,red,->] (.5,.877) arc (60:-20:1);
 \draw[dashed,<-] (1,1.73) arc (60:0:2);
  \end{tikzpicture}
\caption{Schematic view of the glide planes of the [111] zone. A generic MRSS plane is labeled in red, while, by way of example, the $(\bar{1}01)$ is the glide plane. The suffixes `T' and `AT' refer to the \emph{twinning} and \emph{antitwinning} senses, respectively.\label{110}}
\end{figure}

The projection of the external stress tensor on the RSS plane as in eq.\ \ref{s0} is what is known as \emph{Schmid's law}. In the coordinate system depicted in Figure \ref{110}, this results in:
\begin{equation}
\sigma_{ext}=\vec{t}\cdot\matr{\sigma}\cdot\vec{n}=-\sigma_{xz}\sin\theta+\sigma_{yz}\cos\theta
\label{s}
\end{equation}
where the angle $\theta$ is measured from the positive $x$-axis to the glide plane defined by $\vec{n}$. Here, the only active components of the stress tensor that result in a resolved component of the Peach-K\"ohler force on the glide plane are $\sigma_{xz}$ and $\sigma_{yz}$.
In Section \ref{nosh}, we explain how to substitute eq.\ \ref{s} by a suitable projection law that reflects non-Schmid behavior.
In what follows, for brevity, we use the shorthand notation $s$ to denote the stress at any given segment, $s\equiv s_i$, and $\tau$ to refer to the resolved shear stress, $\tau\equiv\sigma_{\rm RSS}$.

In the same spirit as previous works on the topic, our approach is to generate kink-pair configurations by sampling the following general function representing the kink-pair nucleation probability per unit time:  
\begin{eqnarray}
\label{req}
r_i(s;T)=\omega f(s)\exp\left\{-\frac{\Delta H(s)}{kT}\right\}\\
\nonumber f(s)=\left\{\begin{array}{ll}
\frac{l_i-w(s)}{b} & {\rm if}~l_i>w(s) \\
0 & {\rm if}~l_i<w(s)
\end{array}\right.
\end{eqnarray}
where $\omega$ is the attempt frequency, $\Delta H(s)$ is the kink-pair activation enthalpy, $w(s)$ is the kink-pair separation, $k$ is Boltzmann's constant, and $T$ is the absolute temperature. The variable $l_i$ represents the length of a rectilinear screw segment $i$, with $L=\sum_il_i$. Typically, a non screw segment --{\it e.g.}~a kink-- separates each segment $i$ from one another. 

The expression above merits some discussion. The stress-dependent functions $\Delta H(s)$ and $w(s)$ are of the following form:
\begin{equation}
\Delta H(s)=\Delta H_0\left(1-s^p\right)^q\\
\label{heq}
\end{equation}
\begin{equation}
w(s)=w_0(s^{-m}+c)(1-s)^{-n}
\label{weq}
\end{equation}
where $p$, $q$, $w_0$, $m$, $c$, and $n$ are all adjustable parameters. Equation \ref{heq} represents the formation enthalpy of a kink pair at stress $s$ and follows the standard Kocks-Argon-Ashby expression that equals the energy of a pair of isolated kinks at zero stress and vanishes at $s=1$ ($\tau=\sigma_P$) \citep{kocks1975}. For its part, eq.\ \ref{weq} is a phenomenological expressions (no physical basis) that diverges for both limits $s=0$ and $s=1$. This is because the equilibrium kink separation distance is undefined at zero stress, while, at the Peierls stress, the notion of kink pair is itself ill-defined. A physical equation for $w(s)$ could conceivably be obtained by, e.g., generating kink pair configurations within a full elasticity model and measuring the force balance (elastic attraction vs.~stress-induced repulsion) as a function of applied stress. However, as discussed below, kinks display a strong atomistic (inelastic) behavior at short distances and we prefer to obtain its atomistic dependence and fit to a function that captures the divergence for $s=0$ and $s=1$. 

The function $f(s)$ represents the number of possible nucleation sites for a kink pair of width $w$ on a segment $i$ of available length $l_i$. It is through this function that the well-known dependence of the screw dislocation velocity with its length at low stress is introduced.

Kink motion is defined by thermal diffusion at zero stress, characterized by a diffusion coefficient $D_k$, and a stress dependent drift characterized by the following viscous law 
\begin{equation}
v_k=\frac{\sigma_Pb}{B}s
\label{kink}
\end{equation}
where $v_k$ is the kink drift velocity and $B$ is a friction coefficient. Although phonon scattering treatments predict that $B$ increases linearly with temperature, our MD calculations show $B$ to be constant across all temperatures, in agreement with previous studies on kink motion \citep{swinburne2013}. The overall dynamic behavior of kinks must account for both contributions to the mobility, which can be done by treating kink diffusion as a Wiener process within the kMC model in the following fashion. Assuming that a time step $\delta t$ has been selected within the kMC main loop, one can write the incremental position of the kink as:
$$\delta x=v_k\delta t \pm \sqrt{D_k\delta t}$$
where the $\pm$ sign reflects the random character of diffusion. The maximum kink flight time in the code is obtained by inverting the above expression and solving for the parameter $\delta t_{\rm mig}$ with $\delta x=\delta x_{\rm max}$, which is an input parameter to the kMC algorithm (cf.\ Section \ref{impl}).

\subsection{Implementation Details}\label{impl}
\begin{figure}[h]
\includegraphics[width=1.0\linewidth]{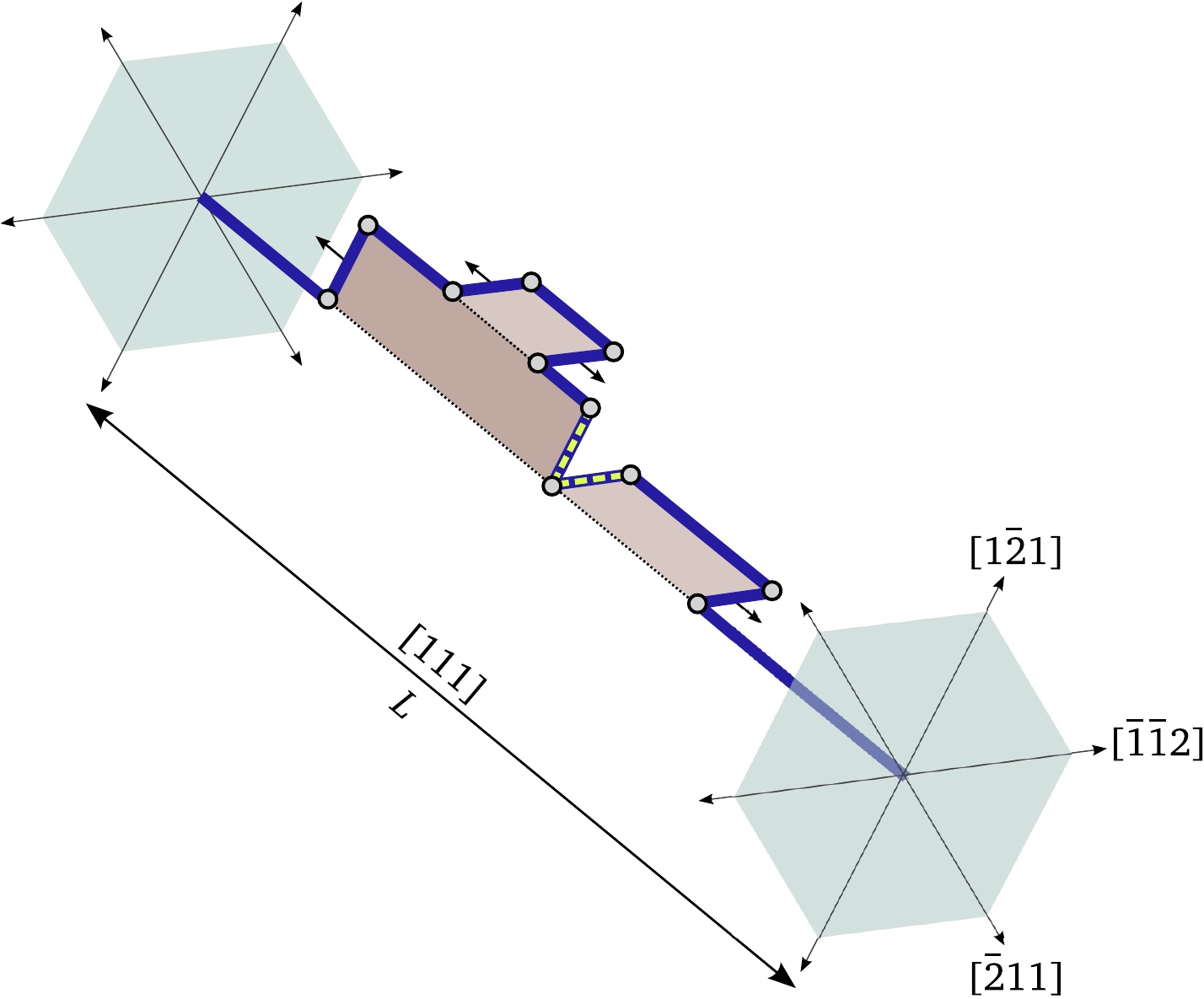}
\caption{Schematic depiction of an arbitrarily kinked screw dislocation line showing kink-pairs on two different $\{110\}$ planes. The arrows indicate the direction of motion of kinks under an applied stress that creates a force on the dislocation in the $[\bar{1}\bar{1}2]$ direction. The dashed line represents a \emph{cross}-kink.\label{schematic}}
\end{figure} 

The dislocation is represented by a piecewise straight line extending a length $L$ 
along the $\nicefrac{1}{2}\left[111\right]$ Burgers vector direction, as depicted in Figure \ref{schematic}.
It consists of pure screw segments, which can be of any length, and
pure edge segments (kinks), which all have the same length $h=\frac{\sqrt{6}}{3}a_{0}$,
the unit kink height. The direction of kink segments can be any one
of six $\nicefrac{1}{3}\left\langle 112\right\rangle $ directions,
corresponding to glide of the dislocation on the three $\left\{ 110\right\} $
planes of the $\left[111\right]$ zone. Periodic boundary conditions
are used in the direction parallel to the screw direction\footnote{Although nothing precludes the use of fixed end points, akin to pinning points in real microstructures.}.

Even though kinks are represented by pure edge segments in our model,
we implicitly assume that kink pairs have a trapezoidal shape of a certain
width $a$ (cf.\ Section \ref{kinke}). This is why the length of a screw segment,
where new kink pairs can nucleate, is effectively reduced by one kink
width $a$. Kink segments move parallel to the $\left[111\right]$
screw direction and can recombine with other kinks of opposite sign.
The local kink-pair nucleation rate (eq. \ref{req}) and the drift velocity
of kink segments (eq. \ref{kink}) depend on the local stress, which is the
superposition of a fixed, externally applied stress tensor and varying internal
stresses (cf.\ eq.\ \ref{s0}). The internal contributions, $\sigma_{ij}$, originating from mutual interactions between the piecewise straight dislocation segments, are computed using non-singular isotropic elasticity
theory with a core width of $0.5b$ \citep{cai2006}. W is a perfectly isotropic elastic material and so using the theory by \citeauthor{cai2006} introduces no limitations in this regard.

The local stress on a given segment $i$ may not be spatially uniform. To resolve this spatial dependence, we sample the local nucleation rate at multiple random positions along $l_i$.
The simulation proceeds in discrete time steps of variable length according
to the following algorithm:
\begin{enumerate}
\item The current drift velocities of existing kinks are computed from the
local stress at the center of each kink.
\item Assuming constant kink velocities, a migration time $\delta t_{\mathrm{mig}}$
is computed, which is the lowest time taken by any kink in the system
to move a prescribed maximum distance $\delta x_{\rm max}=40b$, or before any
kink-kink collision occurs\footnote{We have found that the calculations are quite insensitive to the value of $\delta x_{\rm max}$. By way of example, a fourfold increase or decrease of the nominal value of $40b$ results in only changes of $\approx3\%$ in the kink velocities.}.
\item A nucleation time $\delta t_{\mathrm{nuc}}$ is randomly generated from the
exponential distribution defined by the total nucleation rate, which
is the sum of all kink-pair nucleation rates on all screw segments
and for all kink directions.
\item If $\delta t_{\mathrm{mig}}<\delta t_{\mathrm{nuc}}$, then all kinks move at their
current velocities for a time period $\delta t_{\mathrm{mig}}$ and the simulation
time is incremented accordingly. Otherwise, the kinks move for a time
period $\delta t_{\mathrm{nuc}}$, followed by a kink-pair nucleation on
a screw segment. The nucleation site is chosen according to the local
nucleation rates by a standard kMC algorithm, and the simulation time
is incremented by the reciprocal of the total nucleation rate \citep{voter}.
\item Any kink-kink reactions occurring after the propagation of kinks are
carried out and the topology of the line model is updated. Return
to step 1.
\end{enumerate}
In the last step, kink-kink annihilation and debris dislocation loop
formation is considered. As described by \citet{cai2001} and \citet{marian2004}, two pile-ups of cross
kinks can spontaneously reconnect to form a self-intersection of the
dislocation line. At the self-intersection point, the connectivity
of the line is broken into two independent parts: the
infinite screw dislocation, which continues moving through the material,
and a closed prismatic loop, which remains behind.

Two kinks on the same screw segment, which have formed on different
$\left\{ 110\right\} $ planes, may collide and form a so-called \emph{cross kink}
if their relative velocity is negative. Because they are pushed toward
one another by the local stress, the kinks are thus constrained to move
together with a compound velocity equal to the arithmetic mean of
their respective original velocities.

\section{Fitting the kMC model to atomistic calculations}

\begin{figure}[h]
\centering
\includegraphics[trim=0cm 0.5cm 0cm 1cm, clip=true, width=1.0\linewidth]{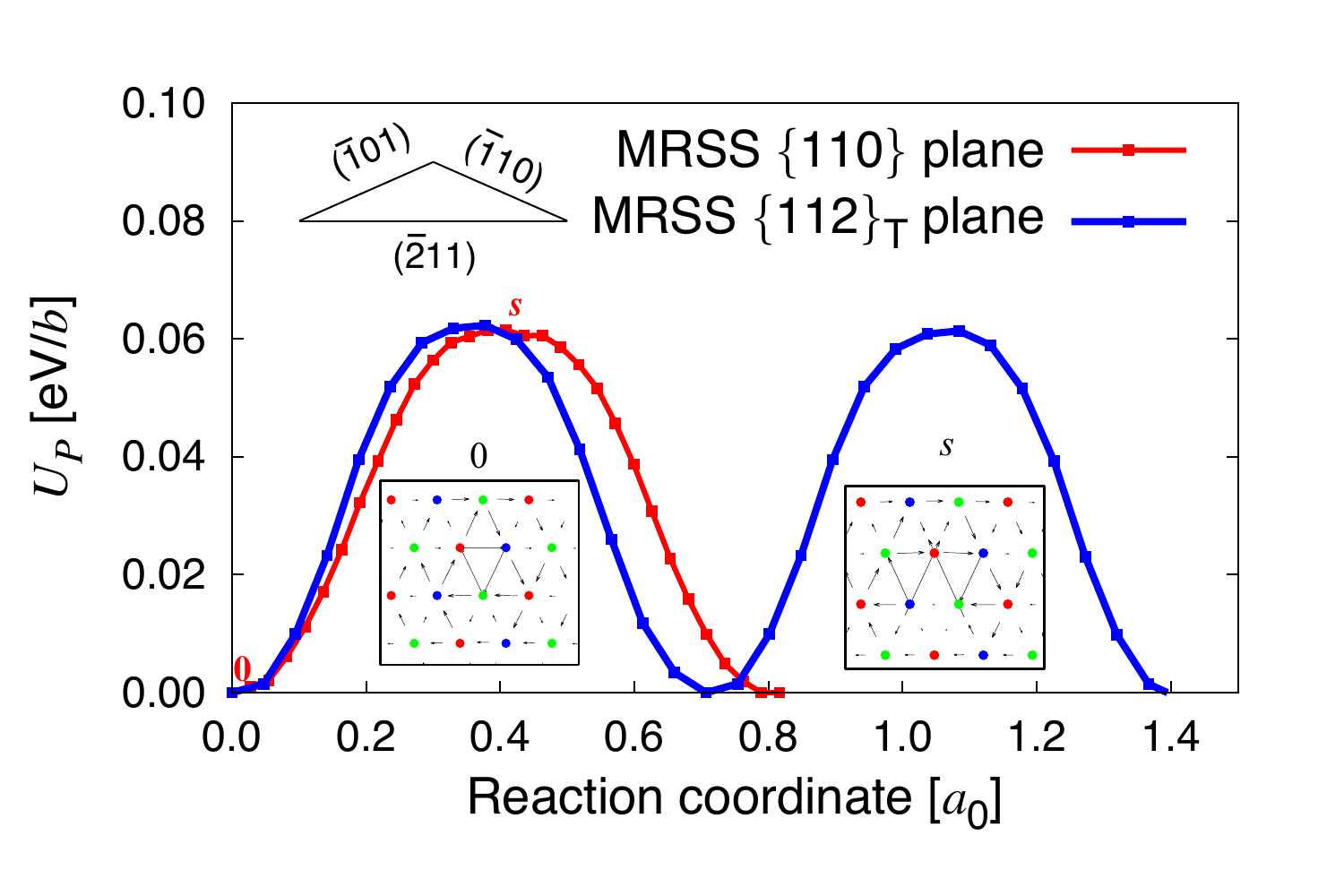}
\caption{Peierls potential for transitions from one equilibrium position to another on a generic $\{110\}$ plane and on a $\{112\}$ plane, twinning sense. Each transition extends over the corresponding reaction coordinate, namely $a_0\sqrt{6}/3$ and $a_0\sqrt{2}$. The geometric decomposition of the $\{112\}$ transition into two alternating $\{110\}$ steps is shown for reference. The two insets show differential displacement maps of the configurations at `0' and `$s$'.\label{joroba}}
\end{figure}

In a previous publication, we have conducted a detailed analysis of several W interatomic potentials for the purpose of screw dislocation calculations \citep{cere2013}. On the basis of that analysis, an embedded-atom method (EAM) \citep{cosmin2013} and a modified EAM (MEAM) potential \citep{park2012} were deemed as the most suitable for screw dislocation property calculations. For reasons of computational efficiency, in this work we choose to perform all supporting calculations for fitting the kMC model with the EAM potential. As a preliminary step, we calculate the Peierls potential on a $\{110\}$ and a $\{112\}$ plane to ascertain whether direct glide on $\{112\}$-type planes is a feasible phenomenon. 
This is done using nudged-elastic band (NEB) calculations of a single screw dislocation in suitably constructed computational cells described below.
The resulting functions represent the substrate potential $U(x)$ as a function of the reaction coordinate $x$ in each case. These are shown in Figure \ref{joroba}, where it is shown that elementary glide on a $\{112\}$ plane is a composite of two elementary steps on alternate $\{110\}$ planes.
Judging by these results, we conclude that glide on any given plane is achieved by way of sequential $\{110\}$ jumps. This is consistent with recent atomistic simulations \citep{gilbert2011,hale2014} and the basis to simulate dislocation glide in the foregoing Sections.

\subsection{Single kink calculations}
\label{singlek}

\subsubsection{Kink energetics}\label{kinke}
Analytical solutions for the kink-pair energy $U_{kp}$ using elasticity models have been proposed by, among others, \citet{dorn1964}, \citet{seeger2002}, and Suzuki and collaborators \citep{koi1993,suzuki1995,eda1996} assuming full elastic and line tension representations of kink-pair configurations and several functional forms for $U(x)$.
However, there is clear evidence in the literature that isolated kink segments display an asymmetry not present in continuum models \citep{mrovec2011,swinburne2013}. This asymmetry emanates from crystallographic and energetic considerations of atomistic nature, and thus calculating kink energies necessitates special methods that capture these particularities. \citet{ventelon2009} have devised a procedure to compute the energies of so-called `left' and `right' kinks, the values of which are given by \citet{cosmin2013} for the current potential:
$$\begin{aligned}
U_{lk}=0.71~\rm{eV}\\
U_{rk}=0.92~\rm{eV}
\end{aligned}$$
The energy of an infinitely separated kink pair is the sum of both energies above: $U_{kp}(\infty)=1.63$ eV.

\begin{figure}[h]
\centering
\includegraphics[width=1.0\linewidth]{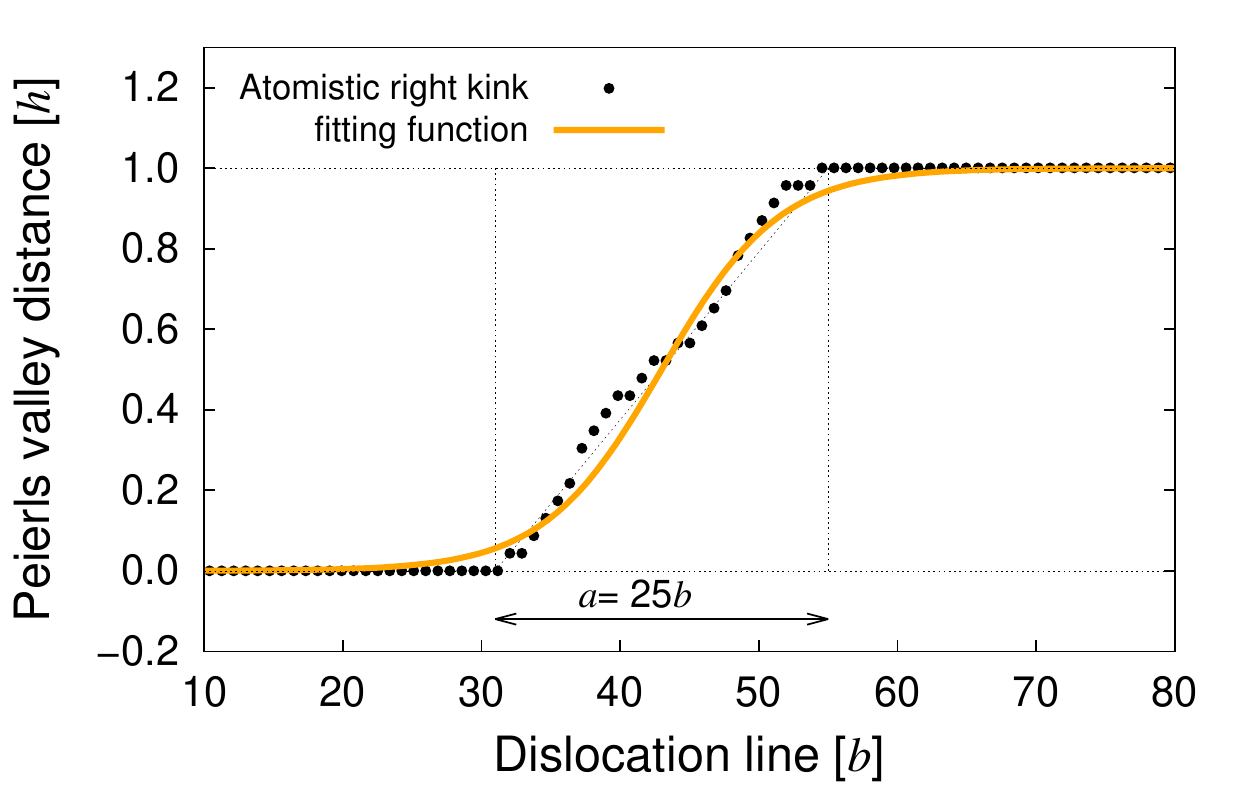}
\caption{Single kink shape as obtained with atomistic calculations. The width of the kink, $a$, is measured from by fitting the data points to a hyperbolic tangent function.\label{isokink}}
\end{figure}

Additional useful information that can be extracted from these calculations is the \emph{width} of an isolated kink, that is, the stretch along the $[111]$ direction over which the kink extends. Figure \ref{isokink} shows the kink shape and its width obtained via Volterra analysis \citep{ventelon2013,gilbert2013}. The kink shape is fit to a function of the form:
$$x(z)=\frac{h}{2}\left(1+\tanh\left(\frac{z}{\epsilon}\right)\right)$$
where $h$ is again the distance between Peierls valleys and $\epsilon$ is a fitting parameter\footnote{Note that here we are using a coordinate system consistent with Fig.\ \ref{110}.}. The kink width $a$ is measured as the distance over which $x(z)$ varies from $0.05h$ to $0.95h$, which is approximately $3\epsilon$. Fitting $x(z)$ to the data points shown in Fig.\ \ref{isokink} yields a value of  $\epsilon=8.4b$ or $a=3\epsilon\approx25b$. This is the value used in the kMC code to represent kinks as trapezoidal elastic segments.

\subsubsection{Kink mobility}

As noted above, kinks can display both mechanical driven (stress-dependent) and diffusive (stress-independent) motion. Both of these must be characterized to define kink motion in the context of the kMC code. In bcc metals, including W, the energy barrier to kink motion on $\{110\}$ planes is negligible. This calls for a diffusion model of the following type:
\begin{equation}
D_k = \frac{kT}{h\gamma_k}
\label{diffkink}
\end{equation}
where $\gamma_k$ is a temperature-independent friction coefficient. For its part, stress-driven drift motion is assumed to follow eq.\ \ref{kink}, which for practical reasons is expressed as:
$$\dot{z} =\frac{\vec{b}\cdot\matr{\sigma}}{B}$$
where $\dot{z}\equiv v_k$.
Atomistic simulations of suitable geometric setups can be performed to obtain $\gamma_k$ by mapping eq.\ \ref{diffkink} to the temperature dependence of the diffusivity, obtained as $D_k = d\langle\Delta z^2\rangle/dt$ with $\langle\Delta z^2\rangle$ the mean square displacement. In turn, $B$ is calculated by obtaining the velocity-stress curves at different temperatures and mapping to eq.\ \ref{kink}, with the two friction coefficients connected through Einstein's relation $B=h\gamma_k$.
The detailed calculations are provided in \ref{a:kink} and are summarized here as well as in Table \ref{table}. An effective diffusivity for left and right kinks is taken:$$D_k(T) = 7.7 \times10^{-10}T$$
with the diffusivity in m$^2$$\cdot$s$^{-1}$ and $T$ in Kelvin. This corresponds to a friction coefficient of $\gamma_k=7.0\times10^{-5}$ Pa$\cdot$s. For the drift velocity we obtain a stress dependence of:
$$v_k =3.8\times10^{-6}\tau$$ 
where the velocity is in m$\cdot$s$^{-1}$ when the stress is given in Pa. This results in a friction coefficient $B=8.3\times10^{-5}$ Pa$\cdot$s. 

\subsection{Kink pair enthalpy}
\label{kp_enth}

\begin{figure}[h]
\centering
\includegraphics[trim=1cm 1.4cm 1cm 1.cm, clip=true,width=\linewidth]{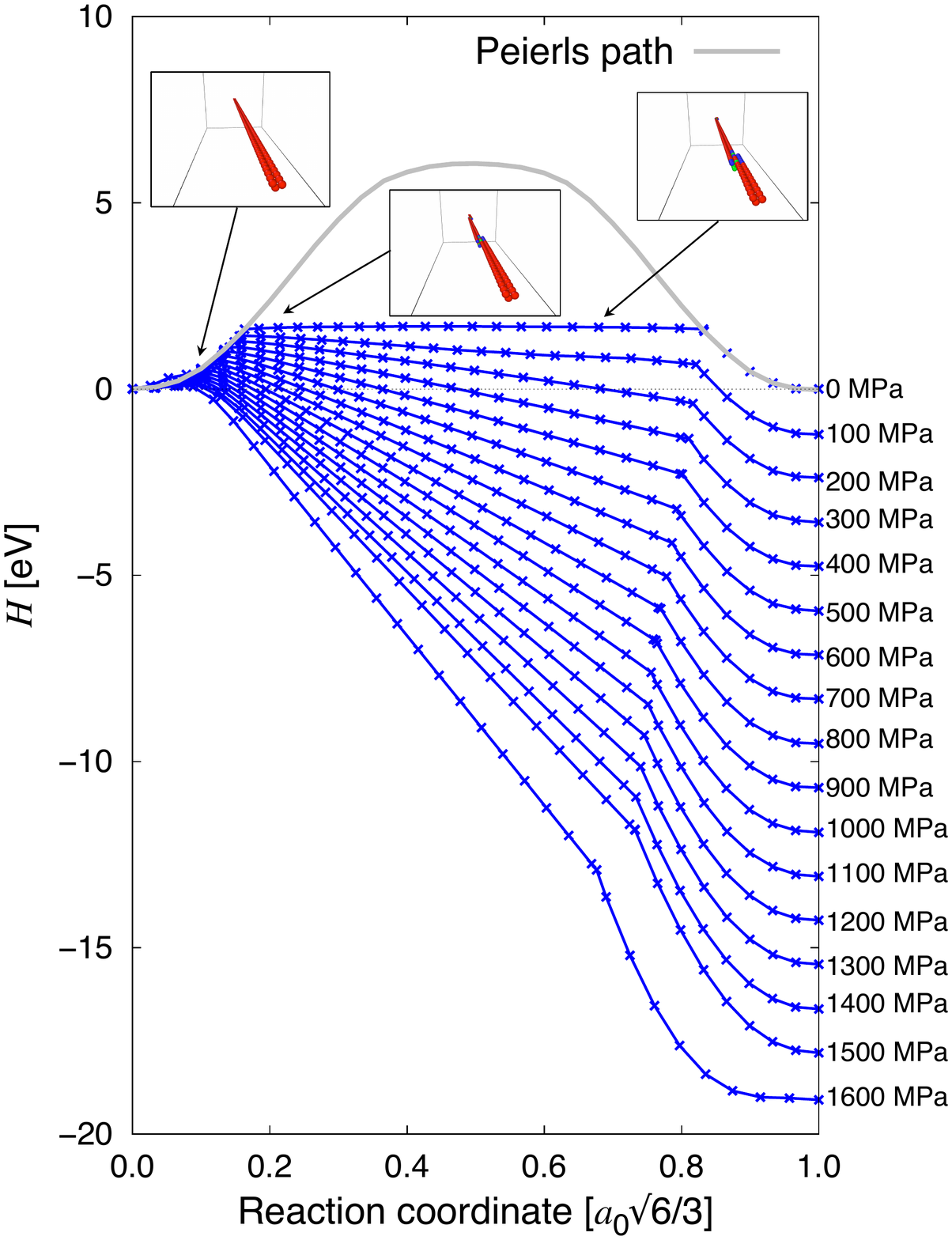}
\caption{NEB transition pathway along the Peierls coordinate for screw dislocation segments of length $100b$. The activated state is taken as the point of maximum enthalpy in each case, where the kink pair separation is measured. The insets correspond to kink pair configurations at several points along the NEB trajectory visualized using the centrosymmetry deviation parameter.\label{NEB}}
\end{figure}

As it was shown in the preceding section, kinks are short dislocation segments displaying a sign asymmetry that cannot be captured by using elasticity theory.  
To compute $\Delta H$, here we take a direct atomistic approach by treating kink pair configurations as activated states of long straight dislocation lines moving along the Peierls trajectory. In the same manner as a number of previous studies \citep{wen2000,rodney2009,gordon2010,raya}, we perform nudged-elastic band (NEB) calculations of screw dislocation lines $100b$ in length going from one Peierls valley to the next as a function of stress. These calculations are periodic along the dislocation line but finite on $\{110\}$ surfaces parallel to the glide plane, where the external shear stress is applied. To break the translational symmetry along the $[111]$ direction, we create intermediate replicas seeded with kink-pair configurations. We then calculate the maximum total energy along the NEB path and measure the kink separation at the activated state. An artifact of this calculations results from using periodic boundary conditions along the line direction for the zero stress case. In these conditions, a separation of exactly 50$b$ is attained, which results in a small but non-negligible elastic interaction energy. Thus, the following limiting values are directly assumed:
$$\Delta H(s=0)=\Delta H_0=U_{kp}=U_{rk}+U_{lk}=1.63~\rm{eV}$$
$$w(s=0)\rightarrow\infty$$ 
Figure \ref{NEB} shows the NEB calculations of the Peierls transition pathway as a function of stress for the screw dislocation lines of length $100b$. The unrelaxed NEB trajectory consists of straight dislocations as the initial and final states, separated by one Peierls valley. The intermediate states are obtained by introducing a kink pair at some arbitrary location along the line, separated by a distance varying linearly from $50b$ ($L/2$) for the second replica to $10b$ for the penultimate one. We then relax the entire trajectory using the nudged elastic band procedure and measure the energy along the path. The final trajectory is obtained as the lowest-energy superposition between the NEB energy path and the Peierls energy for a dislocation of length $L=100b$. The activated state is chosen as the maximum energy point on the final trajectory.

\begin{figure}[h]
\centering
\subfigure[]{
  \includegraphics[width=1.0\linewidth]{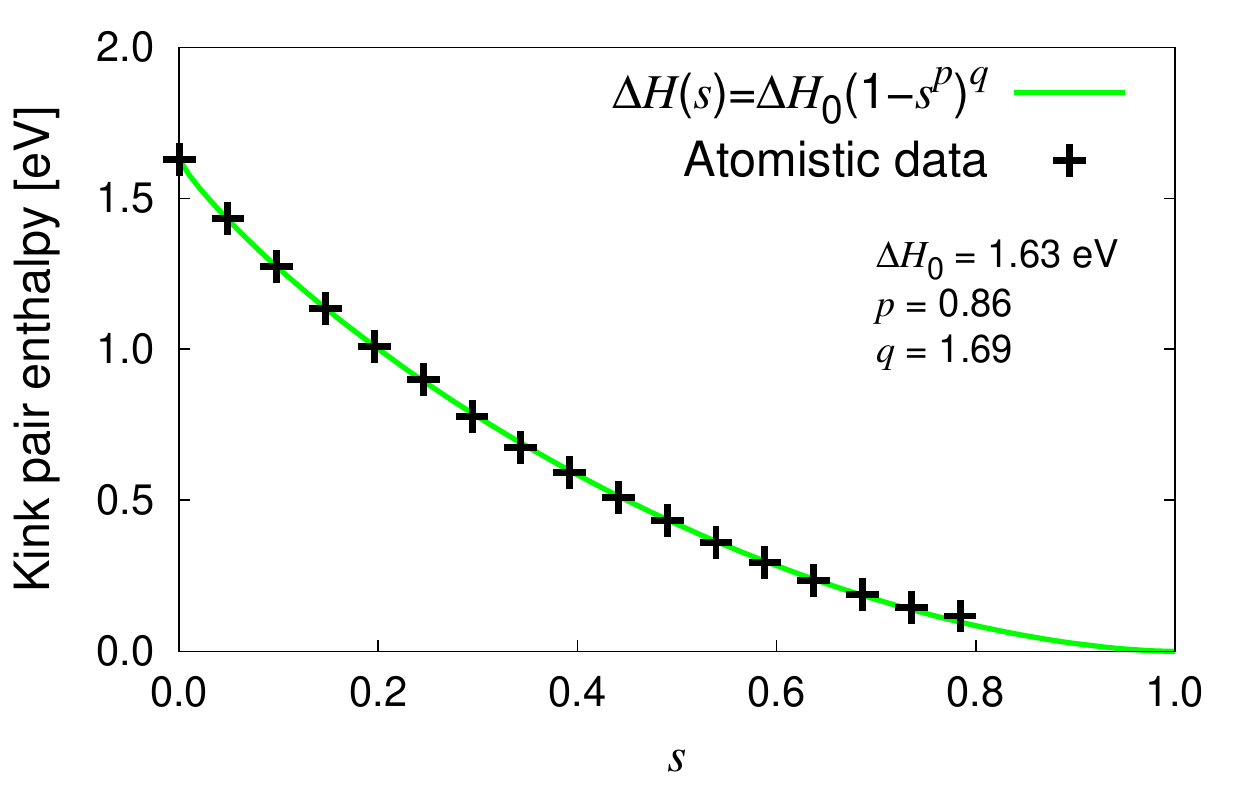}
  \label{subfig_wp}
   }
 \subfigure[]{
  \includegraphics[width=1.0\linewidth]{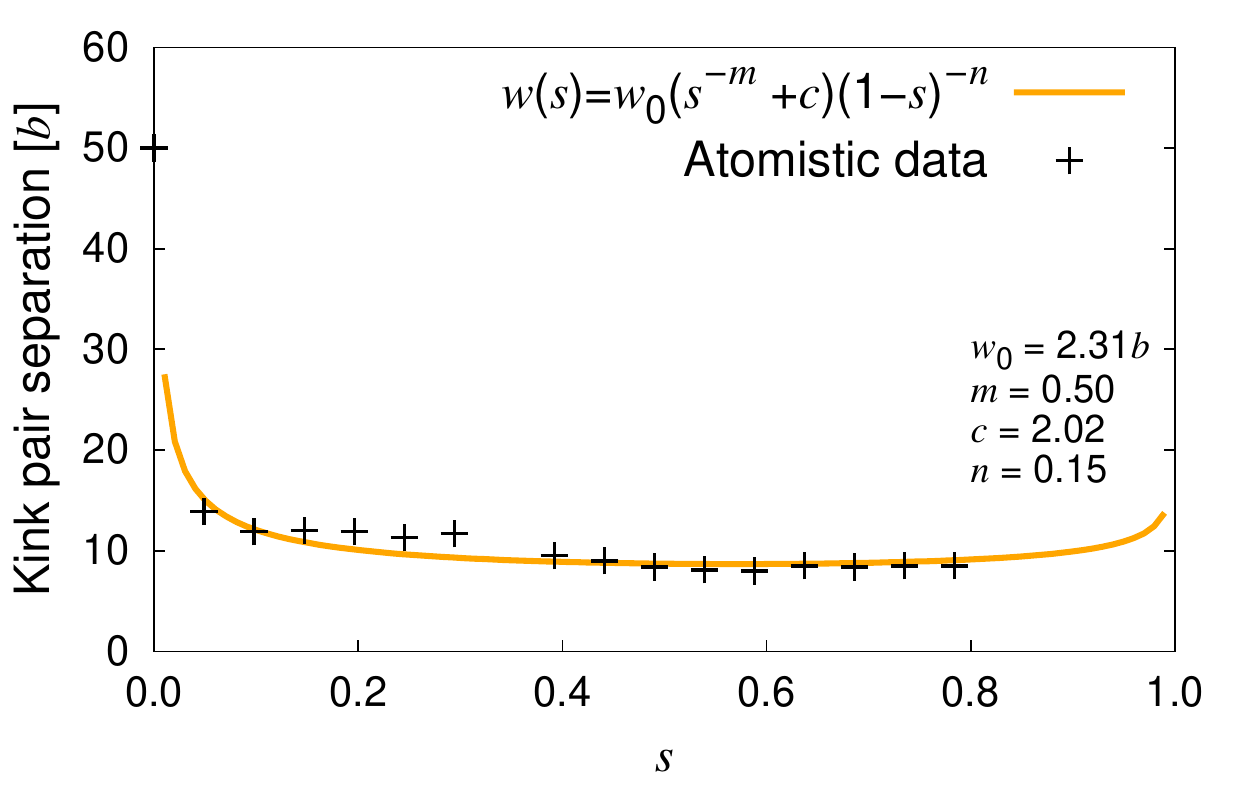}
   \label{subfig_w}
   }
\caption{Kink pair activation enthalpy and separation as a function of stress. The data in each case are fitted to eqs.\ \ref{heq} and \ref{weq}, with the resulting fitting parameters shown in each case.\label{crook}}
\end{figure}

Figures \ref{subfig_wp} and \ref{subfig_w} show the extracted activation enthalpies and separation distances as a function of stress. Fits to eqs.\ \ref{heq} and \ref{weq} result in the parameters given in Table \ref{table}, which are then used in eq.\ \ref{req} for the kMC simulations.

\subsection{Attempt frequency}

The attempt frequency $\omega$ is chosen to be the fundamental mode of the Granato-L\"ucke vibrating string model \citep{lin1999}:
\begin{equation}
\omega=\frac{\pi C_t}{\lambda}
\end{equation}
where $C_t$ is the shear wave velocity and $\lambda$ is a characteristic wavelength. For the purpose of this paper, $C_t$ can be obtained as:
$$C_t=\sqrt{\frac{\mu}{\rho}}$$
where $\mu$ is the shear modulus and $\rho$ is mass density of W. $\rho$ can be trivially obtained from the inverse of the atomic volume $\Omega=a_0^3/2$.
The parameter $\lambda$ is the wavelength of the vibrating undulation, which in this case can be taken as $\lambda=w+a$. Using the parameter values listed in Table \ref{table} and, from Fig.\ \ref{subfig_w}, an effective kink pair separation of $w=11b$, we obtain $\omega=9.1\times10^{11}$ s$^{-1}$.

\subsection{Non-Schmid law from atomistic calculations}
\label{nosh}
Schmid's law states that a slip system will become activated when shear stress, resolved on the slip plane and in the slip direction, reaches a certain critical value called critical resolved shear stress (CRSS). This implies (i) that the CRSS does not depend on the orientation of the load axis, and (ii) that the CRSS is independent of the sign of the loading direction (tension or compression). Many authors have now demonstrated, first, that in bcc crystals the loading symmetry is broken, and, second, that there is a coupling between CRSS and non-glide stress components, all resulting in a breakdown of Schmid's law \citep{duesbery1998,ito2001,rao2001,groger2005,chen2013,barvinschi}. 

Here, our approach is to study deviations from Schmid behavior solely when pure shear stress is applied on different maximum resolved shear stress  (MRSS) planes. We use the standard geometry of the $[111]$ zone as shown in Fig.\ \ref{110} to compute the CRSS using atomistic simulations.
The CRSS is calculated as a function of the angle $\chi$ between the primary glide plane and the MRSS plane. For simplicity, in the atomistic calculations the primary glide plane is represented by $\theta=0$ (cf.\ Fig.\ \ref{110}) and, then, by symmetry, only the angular interval $-\frac{\pi}{6}<\chi<+\frac{\pi}{6}$ need be explored.

\begin{figure}[h]
\centering
\fbox{\includegraphics[width=1.0\linewidth]{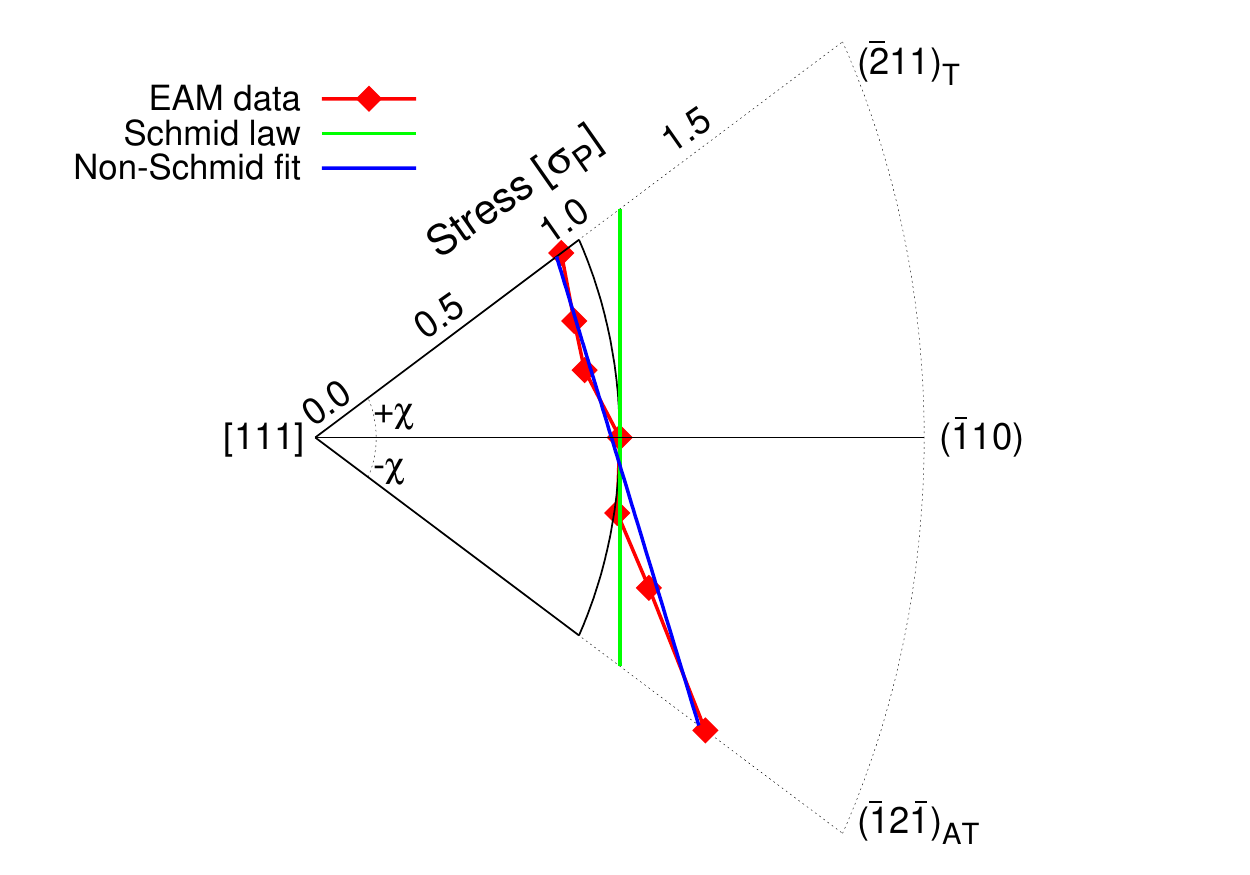}}
\caption{Dependence of the critical resolved shear stress with the angle between the MRSS plane and the primary glide plane. The standard Schmid law is shown as a vertical green-colored line.\label{fig:schmid}}
\end{figure}

The calculations are done by performing atomic relaxations of a single screw dislocations in crystals with periodic boundary conditions subjected to various levels of applied stress. The size of the simulation box is $1\times21\times24$ multiples of the bcc lattice vectors $[111]\times[\bar{1}2\bar{1}]\times[\bar{1}01]$ containing nominally 3024 atoms.
This setup is essentially identical to that used in other atomistic studies. 
The dependence of the CRSS with $\chi$ for the EAM potential employed here is given in Figure \ref{fig:schmid}.
The figure also shows a fit to the data according to the expression:
$$\sigma_c^\chi=\frac{a_1\sigma_P}{\cos\chi+a_2\cos\left(\pi/3+\chi\right)}$$
which is customarily used to represent deviations from the Schmid law \citep{vitek2004,chaussidon2006}. A least-squares fit to the data yields $a_1=1.26$ and $a_2=0.60$, which are added to Table \ref{table}. The details about the implementation of this equation into the kMC code for simulations of non-Schmid glide are given in \ref{a:schmid}.

\begin{table}[h]
\caption{List of parameters and functional dependences for fitting the kMC model. All of these parameters have been obtained using dedicated atomistic calculations.}
\centering
\begin{tabular}{llc}
\hline
{\small parameter} & {\small value or function} & {\small units} \\
\hline\vspace{-3mm}\\
$a_0$ & 3.143 & \AA\\
$h$ & $a_0\sqrt{6}/3$ & \AA\\
$\mu$ & 161 & GPa \\
$\nu$ & 0.28 & - \\
$\omega$ & $9.1\times10^{11}$ & s$^{-1}$ \\
$\sigma_P$ & 2.03 & GPa \\
$a$ & $25$ & $b$ \\
$v_k$ & $\tau b/B$ & m$\cdot$s$^{-1}$ \\
$B$ & $8.3\times10^{-5}$ & Pa$\cdot$s \\
$D_k$ & $kT/h\gamma_k$ & m$^2$$\cdot$s$^{-1}$ \\
$\gamma_k$ & $7.0\times10^{-5}$ & Pa$\cdot$s \\
$\Delta H(s;T)$ & $\Delta H_0\left(1-s^p\right)^q$ & eV \\
$\Delta H_0$ & 1.63 & eV \\ 
$p$ & 0.86 & - \\
$q$ & 1.69 & - \\
$w(s)$ & $w_0(s^{-m}+c)(1-s)^{-n}$ & $b$ \\
$w_0$ & 2.31 & $b$ \\
$c$ & 2.02 & - \\
$m$ & 0.50 & - \\
$n$ & 0.15 & - \\
$\sigma_c^\chi$ & $\frac{a_1\sigma_P}{\cos\chi+a_2\cos\left(\pi/3+\chi\right)}$ & GPa 
\vspace{2mm}\\
$a_1$ & 1.26 & - \\
$a_2$ & 0.60 & - \\
\hline
\end{tabular}
\label{table}
\end{table}

\section{Results}

In this section, we calculate the dislocation velocity for a number of different conditions. The velocity is obtained as the derivative of the average position of the dislocation projected on the MRSS plane with respect to time.
We study loading on both $\{110\}$ and $\{112\}$ MRSS planes at different temperatures and stresses. We also investigate three different initial dislocation line lengths: $100b$ is near the maximum extent of what can be presently simulated in MD simulations; $1000b$ is near the average dislocation segment length ($L\approx\rho_d^{-\nicefrac{1}{2}}$) in well-annealed W single crystals \citep{wbook1999}, and $4000b$ is approximately one micron in length. We study stresses from zero to just below the Peierls stress $0<\sigma_{MRSS}<0.9\sigma_P$ and temperatures from room temperature to 1800 K in 300-K intervals. The stress interval ensures that thermal activation is the operating dynamic mechanism, while the temperature limits are roughly those where severe embrittlement and recrystallization are known to limit the usefulness of W as a structural material \citep{wbook1999}.

\subsection{Numerical calculations}
\begin{figure}[h]
\centering
\includegraphics[width=\linewidth]{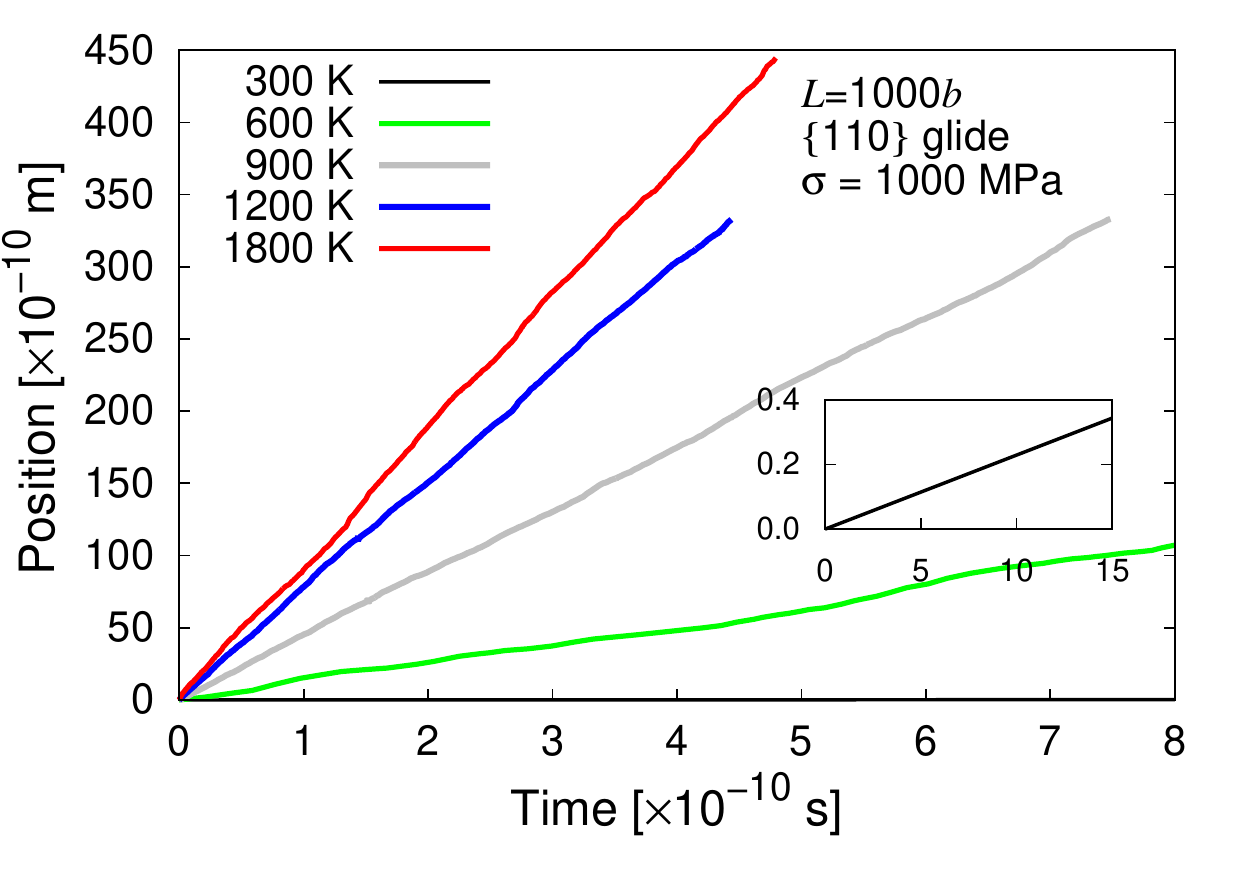}
\caption{Average dislocation position vs.~time at different temperatures for $L=1000b$ with a RSS of 1000 MPa on a $\{110\}$ plane. These simulations were carried out including non-Schmid effects. Due to differences in scale, the trajectory at 300 K is shown separately in the inset.\label{linear}}
\end{figure}
Figure \ref{linear} shows an example of the position vs.~time curves for a dislocation of length $1000b$ at different temperatures and an applied stress of 1000 MPa on a $\{110\}$ plane. Velocities are obtained from linear fits to the data. All simulations share the same qualitative features as those shown in the figure. This linear behavior has been confirmed at room temperature and low stresses in carefully-performed experiments in Fe \citep{caillard2010,caillard2013}.

\begin{figure}[h]
\centering
\subfigure[$L=100b$, $\{110\}$ loading]{
  \includegraphics[width=1.0\linewidth]{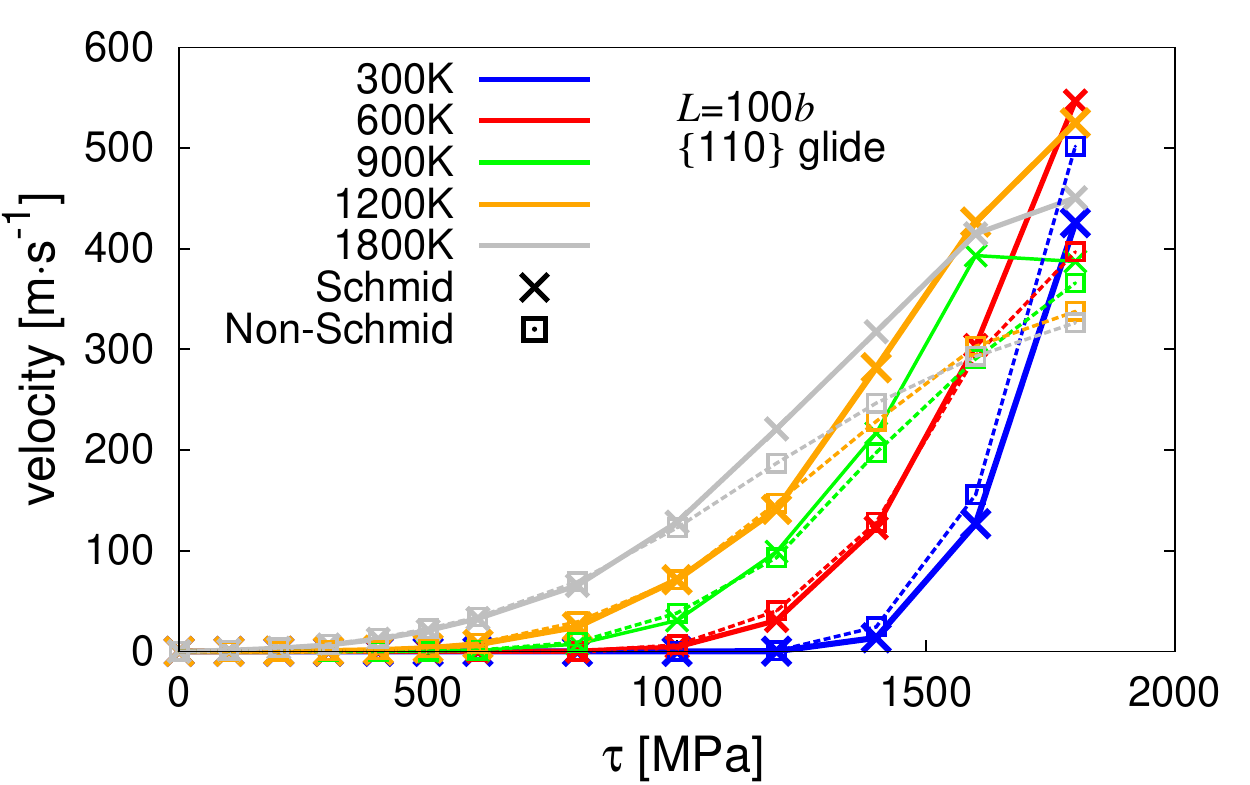}
   \label{subfig1}
   }
 \subfigure[$L=100b$, $\{112\}$ loading]{
  \includegraphics[width=1.0\linewidth]{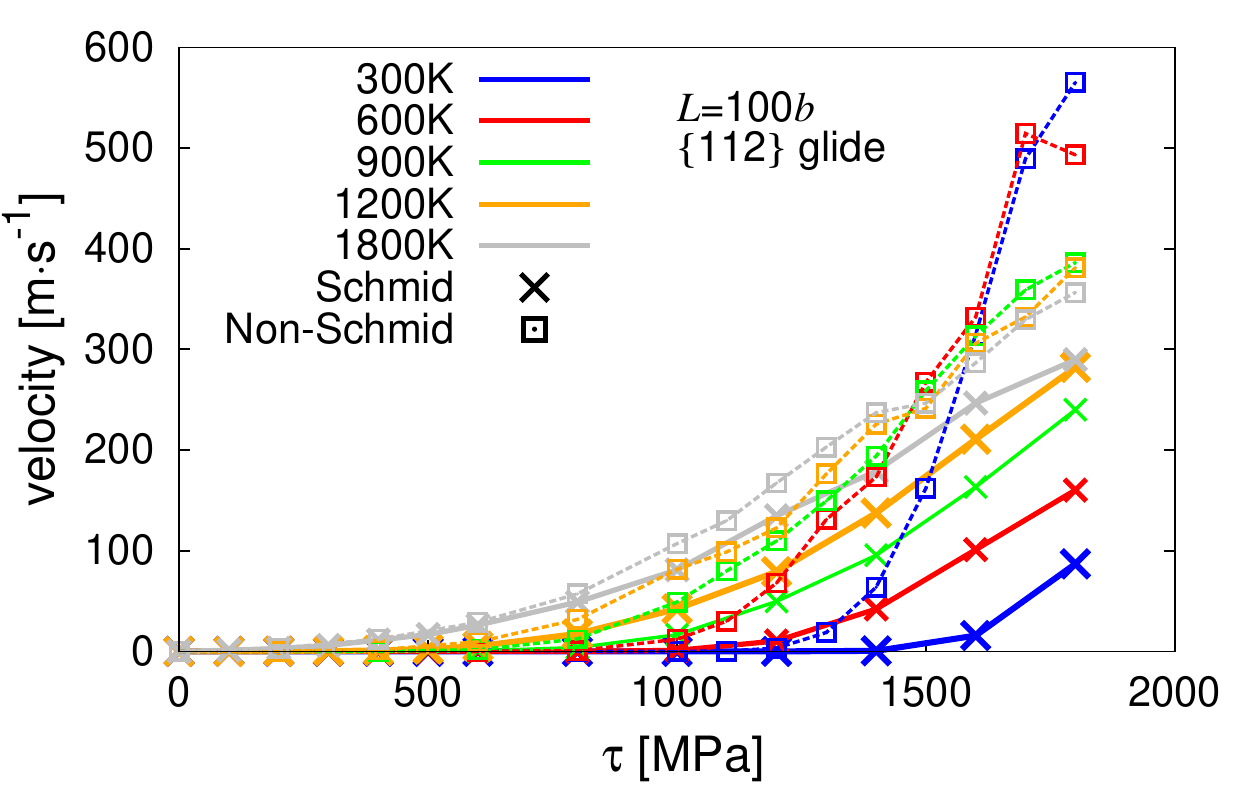}
   \label{subfig2}
   }
\caption{Velocity-stress relations for $L=100b$ for all temperatures, stresses, and including Schmid and non-Schmid loading.\label{100b}}
\end{figure}

\begin{figure}[h]
\centering
\subfigure[$L=1000b$, $\{110\}$ loading]{
  \includegraphics[width=1.0\linewidth]{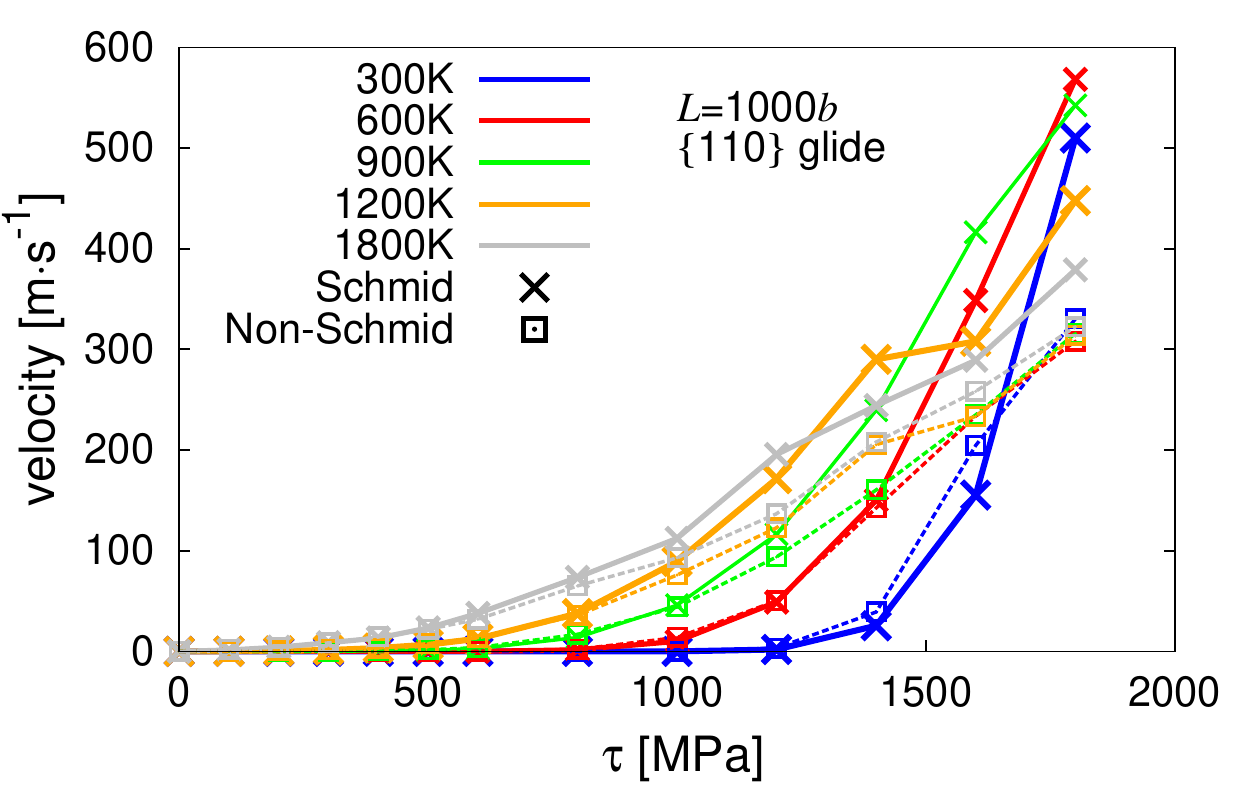}
   \label{subfig3}
   }
 \subfigure[$L=1000b$, $\{112\}$ loading]{
  \includegraphics[width=1.0\linewidth]{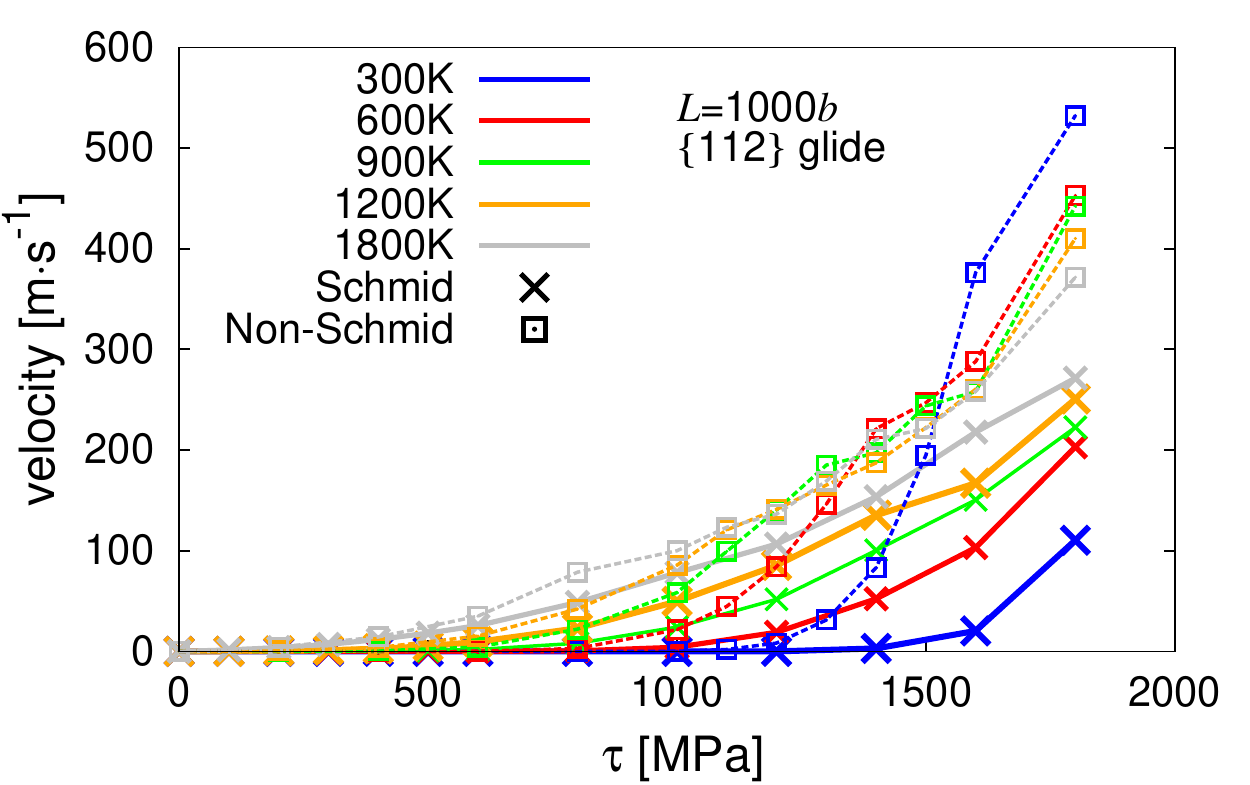}
   \label{subfig4}
   }
\caption{Velocity-stress relations for $L=1000b$ for all temperatures, stresses, and including Schmid and non-Schmid loading.\label{1000b}}
\end{figure}

\begin{figure}[h]
\centering
\subfigure[$L=4000b$, $\{110\}$ loading]{
  \includegraphics[width=1.0\linewidth]{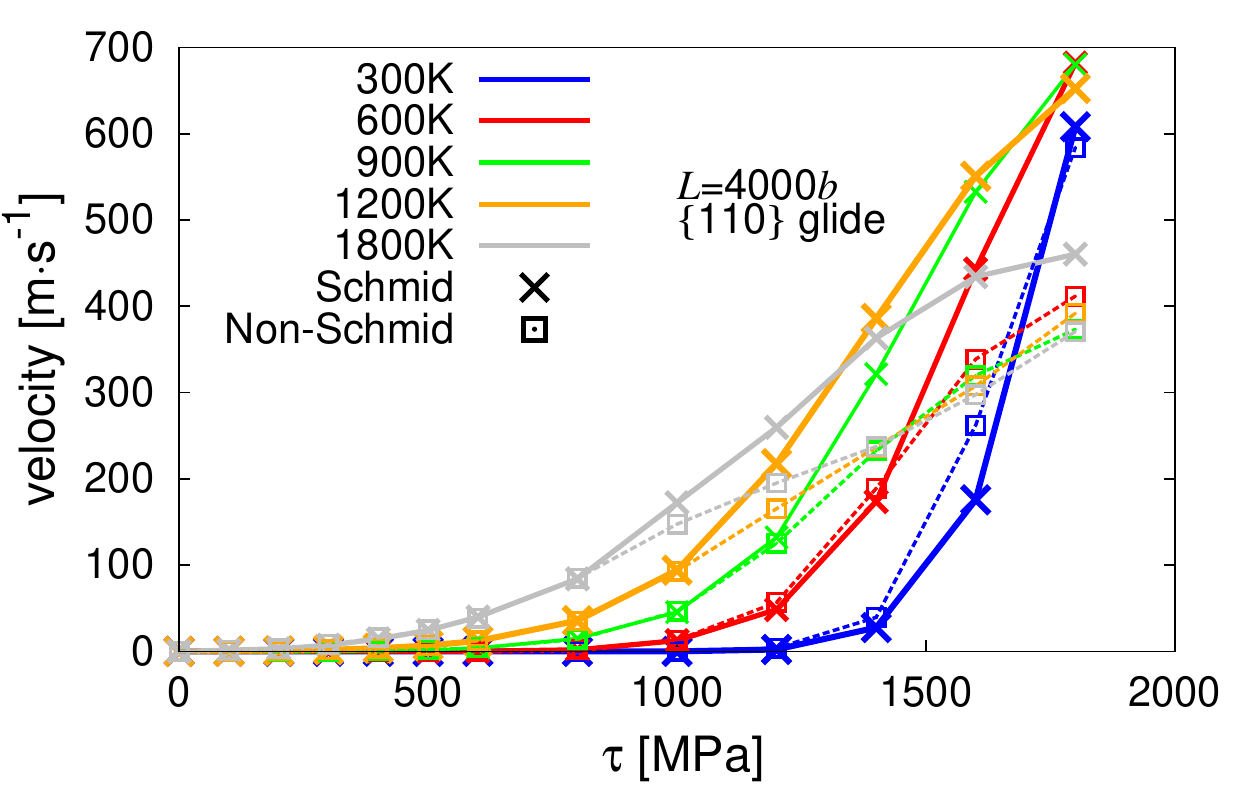}
   \label{subfig5}
   }
 \subfigure[$L=4000b$, $\{112\}$ loading]{
  \includegraphics[width=1.0\linewidth]{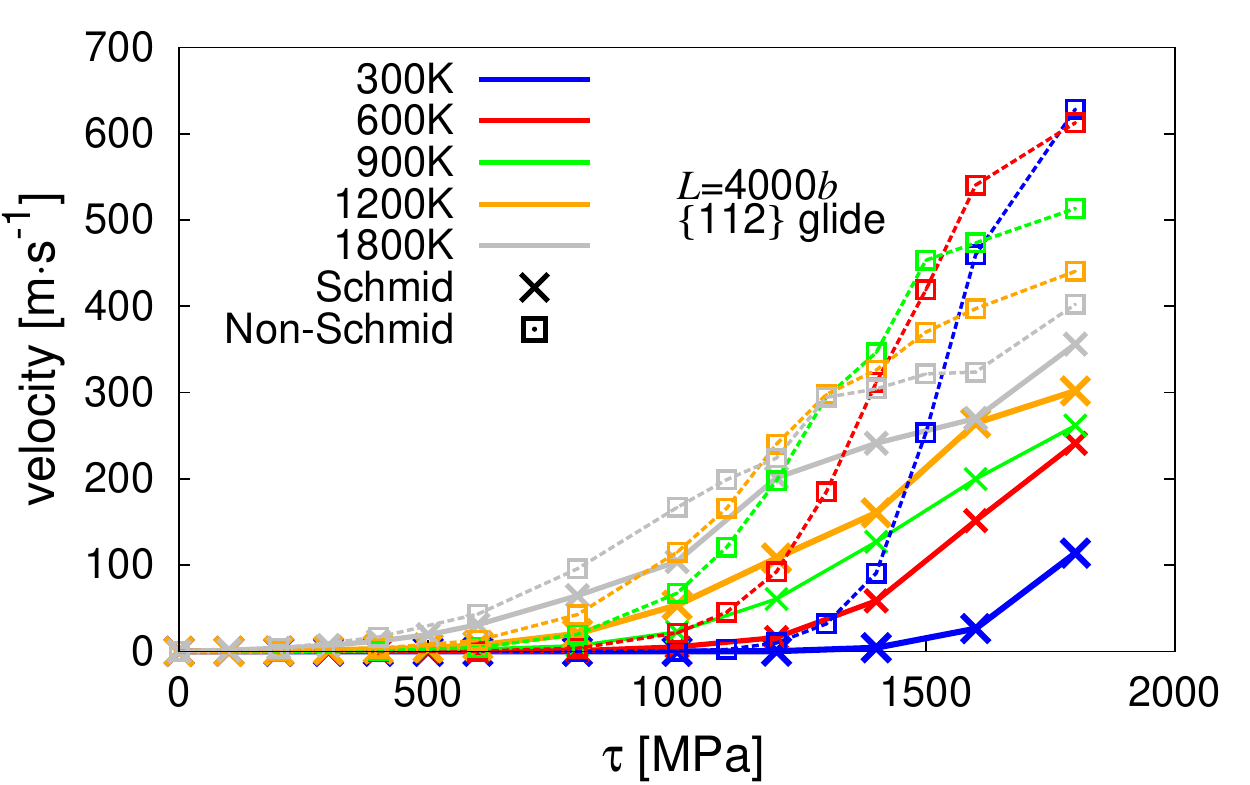}
   \label{subfig6}
   }
\caption{Velocity-stress relations for $L=4000b$ for all temperatures, stresses, and including Schmid and non-Schmid loading.\label{4000b}}
\end{figure}

In Figs.\ \ref{100b}, \ref{1000b}, and \ref{4000b} we provide detailed results as a function of $\tau$ and $T$ for each value of $L$. Each panel includes velocities considering Schmid and non-Schmid effects. 
In advance of discussing these results and their implications in detail in the following section, we note the following features from the figures:
\begin{enumerate}
\item When the MRSS is applied on $\{110\}$ planes, using Schmid's law results in velocities that are larger than when considering non-Schmid effects. This difference is negligible at stresses below 600 MPa and gradually grows up to a factor of two in some cases.
\item On $\{112\}$ planes, by contrast, this tendency is reversed, with the difference being noticeable already at low stresses.
\item When including non-Schmid effects, dislocations move faster at lower temperatures than at higher ones at the highest stresses (>1400 MPa). We will show below that this is a consequence of self-pinning, which is favored in that regime. Using Schmid's law, this tendency is observed in some selected cases, but not generally.
\item At high stresses, there are no appreciable differences between the velocity response for $L=100b$, $L=1000b$, and $L=4000b$ lines. A detailed investigation of the length dependence of the dislocation velocity will be conducted below.
\end{enumerate}

\subsection{Dislocation length dependence}
\begin{figure}[h]
\centering
\includegraphics[width=\linewidth]{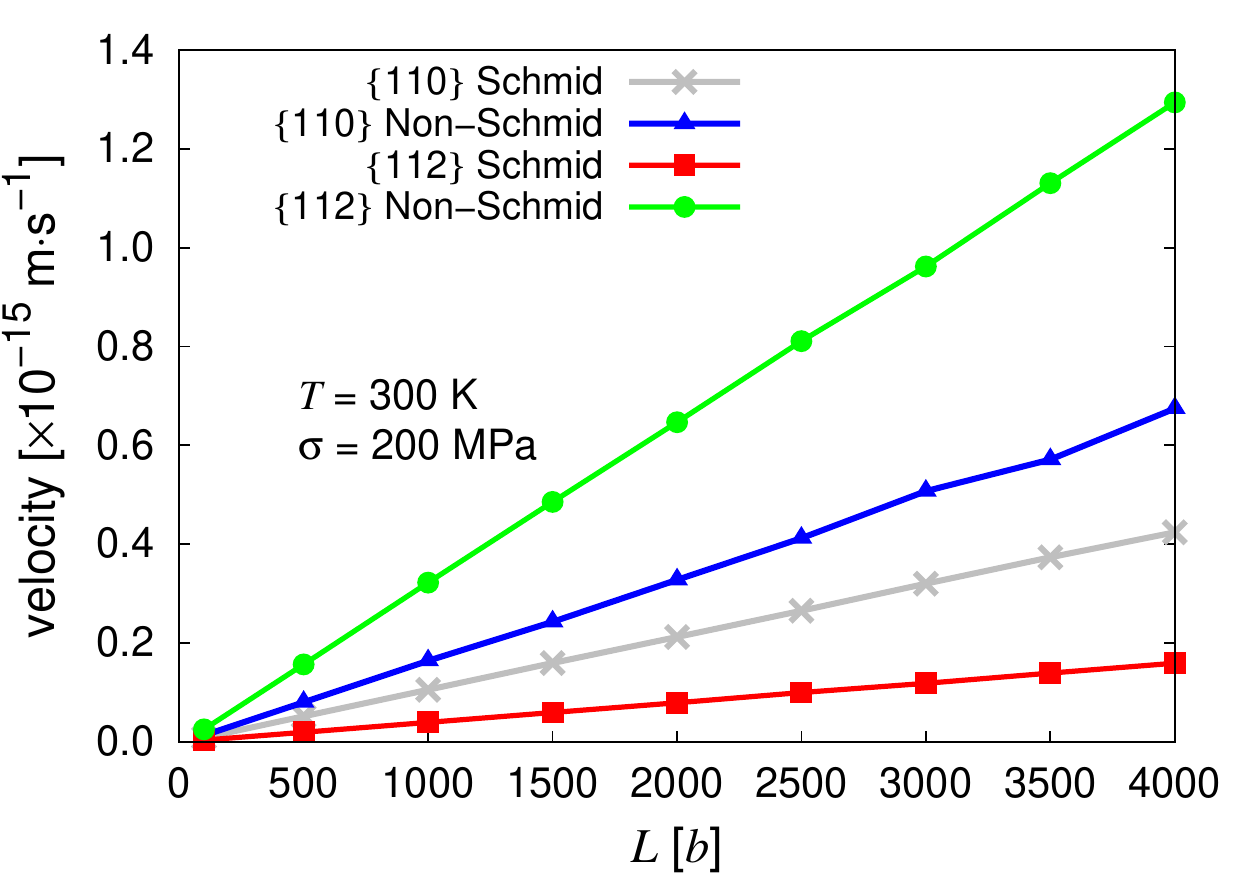}
\caption{Dependence of the dislocation velocity on its initial length for a resolved shear stress of 200 MPa at 300 K and under Schmid and non-Schmid conditions.\label{length1}}
\end{figure}
It has been traditionally assumed that screw dislocation velocity depends linearly on its length, a dependence introduced by construction in dislocation dynamics models \citep{tang1998,naamane2010} but also confirmed experimentally in some limited cases at room temperatures and low stresses \citep{caillard2010,caillard2013}. Here we perform a systematic study of dislocation velocity as a function of $L$ at several temperatures and stresses, and for Schmid and non-Schmid conditions. First we study nominally the same regime as in the experimental works by \citeauthor{caillard2010}, {\it i.e.}~room temperature (300 K) and low stress (200 MPa). We present results for the two slip systems of interest in Figure \ref{length1}, where the linear dependency is clearly distinguished. This is a direct consequence of the form of eq.\ \ref{req}, when nucleation is the rate-limiting step and the dynamics is governed by the existence of one single kink-pair on the line at a given time, {\it i.e.}~$l_i\equiv L$. This is the expected behavior at low strain rates or in quasistastic conditions.

\begin{figure}[h]
\centering
\subfigure[$\tau=1000$ MPa, $\{110\}$ loading]{
  \includegraphics[width=0.85\linewidth]{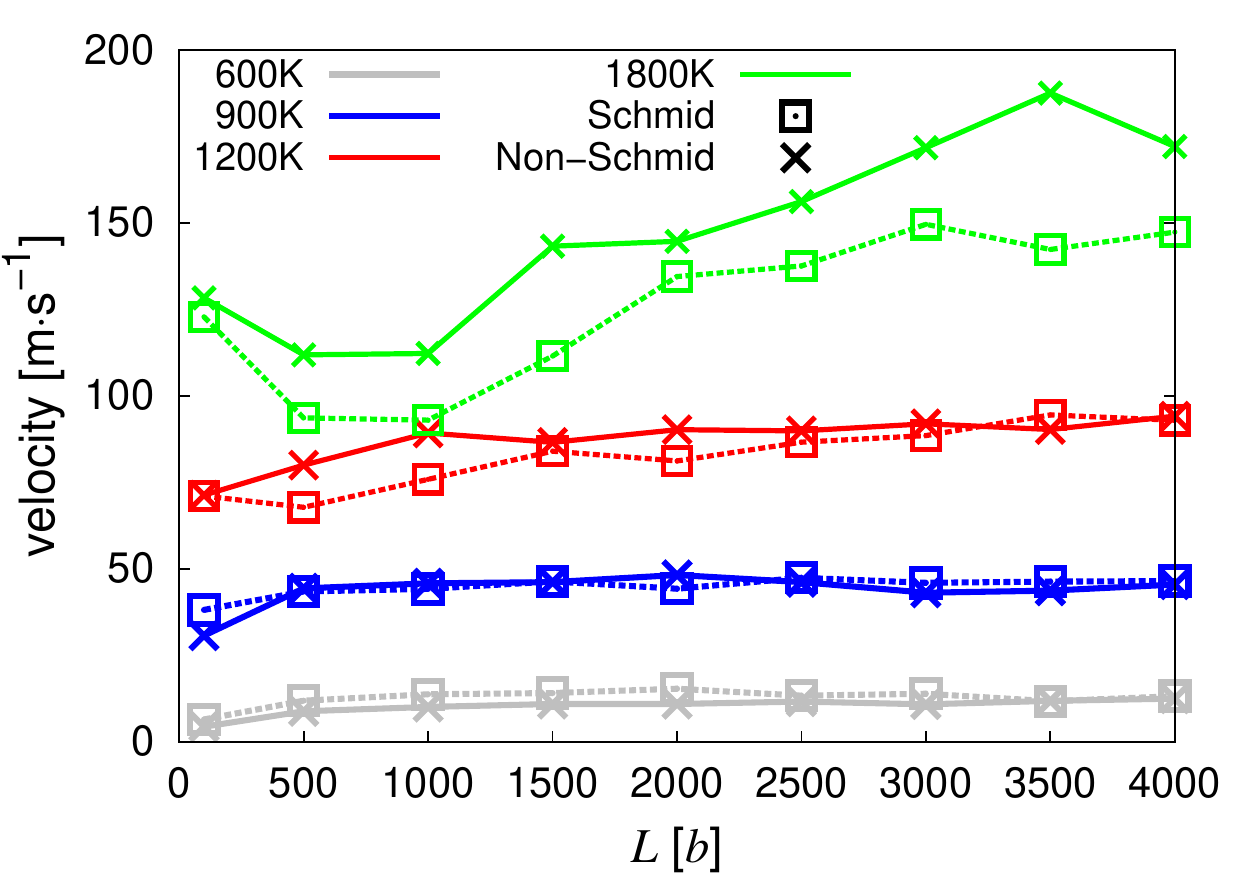}
   \label{subfig7}
   }
 \subfigure[$\tau=1000$ MPa, $\{112\}$ loading]{
  \includegraphics[width=0.85\linewidth]{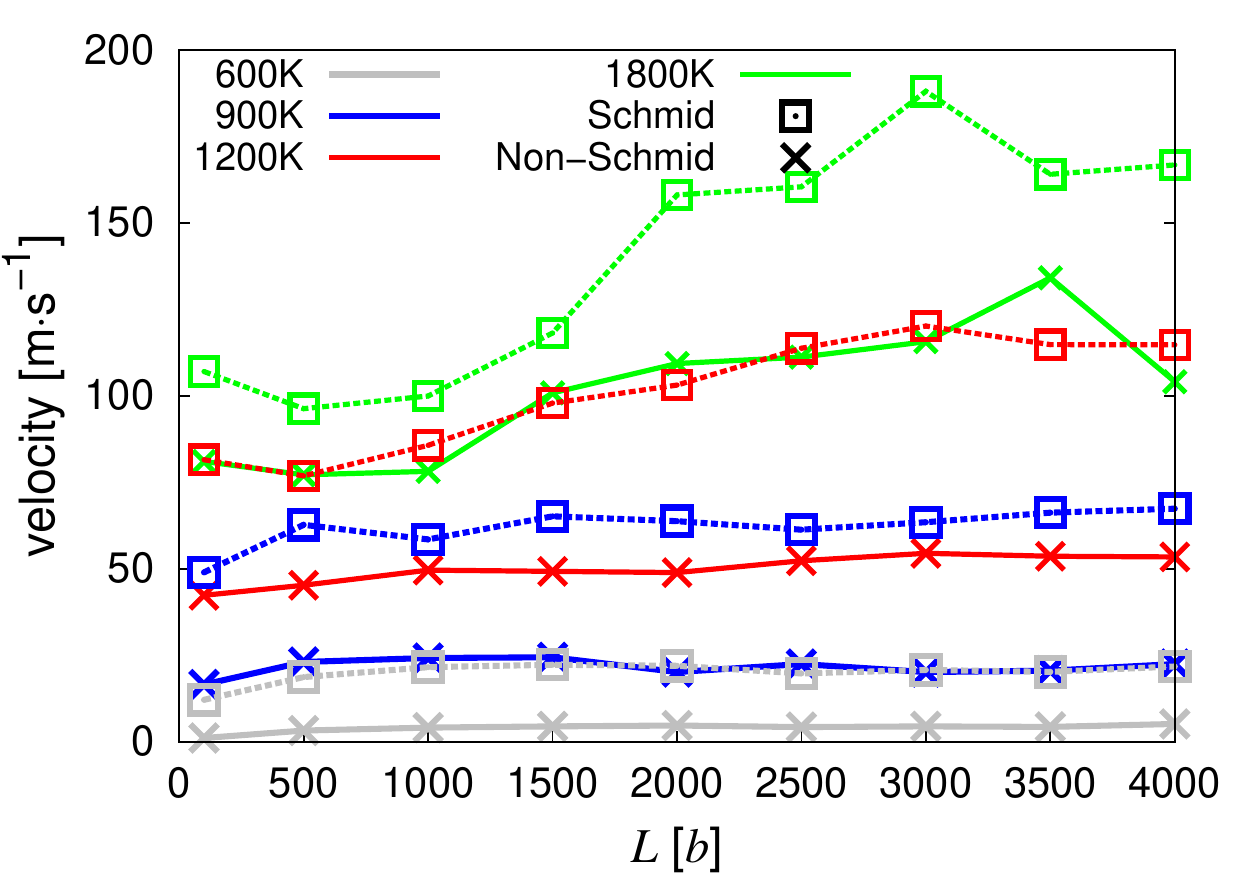}
   \label{subfig8}
   }
\caption{Dependence of the dislocation velocity on its initial length for a resolved shear stress of 1000 MPa at high temperatures and under Schmid and non-Schmid conditions.\label{length2}}
\end{figure}

However, as the stress and/or the temperature increase, this trend becomes gradually weakened until it is lost altogether. Figure \ref{length2} shows results for $\tau=1000$ MPa at different temperatures. In this situation, multiple kink-pairs may coexist at once, giving rise to cross-kinks and other self-pinning features that remove the linear dependence on $L$. These are the conditions that are representative of high-strain rate situations.

\subsection{Trajectory}

\begin{figure}[h]
\centering
\subfigure[$T=300$ K and $\tau=200$ MPa.]{
  \includegraphics[width=1.0\linewidth]{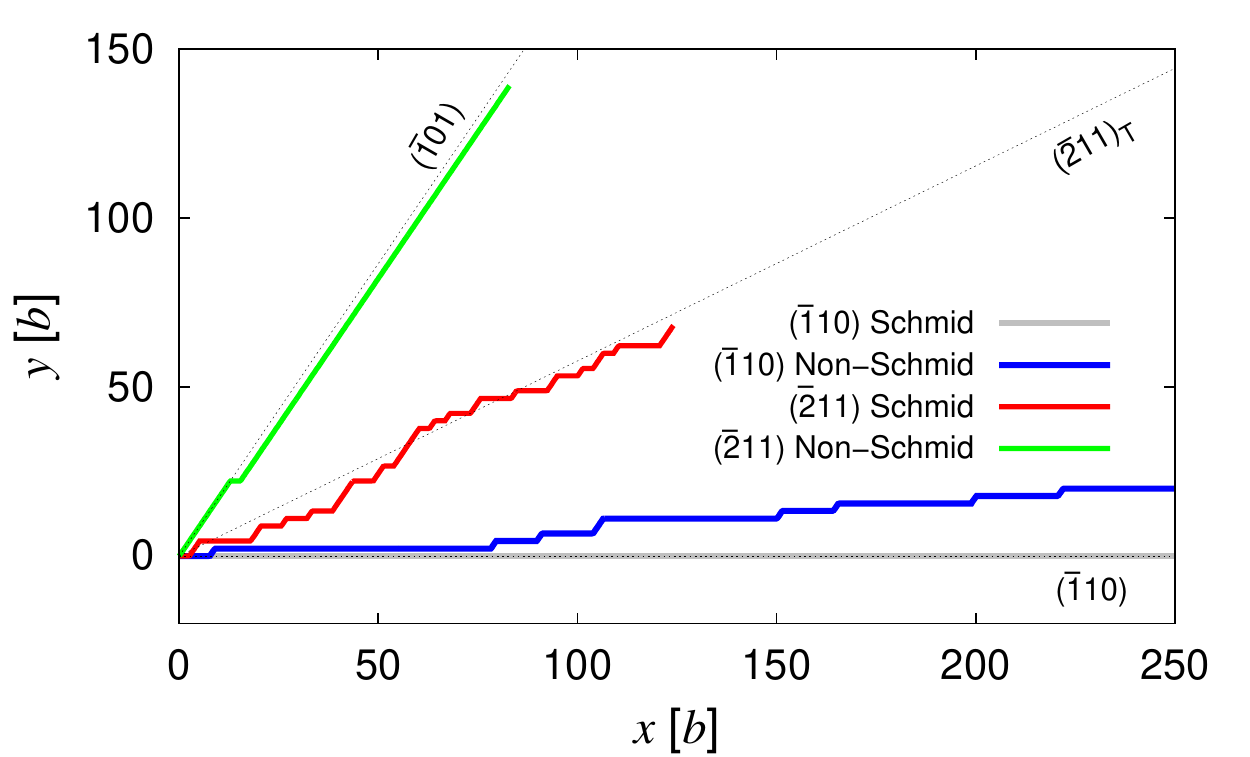}
   \label{traj1}
   }
 \subfigure[$T=1800$ K and $\tau=1000$ MPa.]{
  \includegraphics[width=1.0\linewidth]{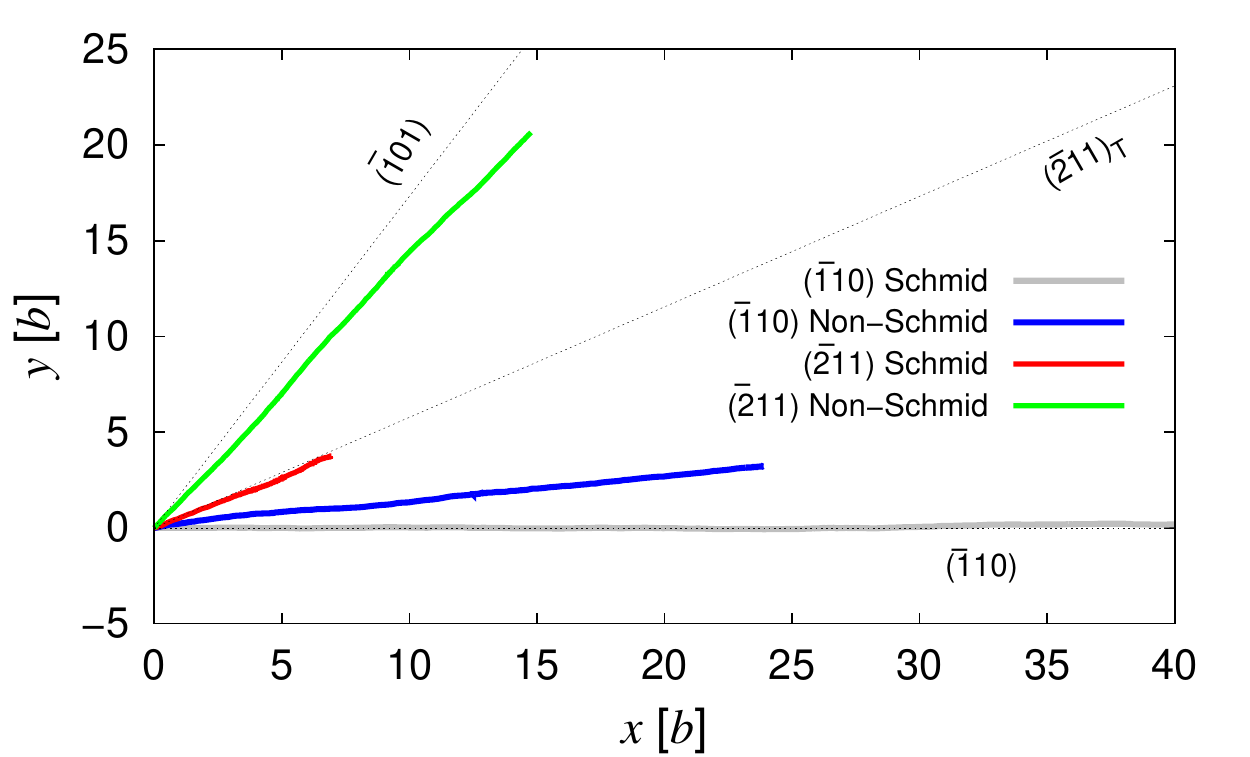}
   \label{traj2}
   }
\caption{Trajectory of a dislocation line of length $L=4000b$ under Schmid and non-Schmid conditions.\label{traj}}
\end{figure}

Next we analyze the impact of considering non-Schmid effects on the trajectory of a screw dislocation projected on the [111] plane. Figure \ref{traj1} shows an example at 300 K and 200 MPa where the MRSS is on the $(\bar{1}10)$ plane (for this analysis we use the axes convention given in Fig.\ \ref{110}). As the figure shows, considering non-Schmid effects results only in a slight deviation from the MRSS plane, characterized by sporadic slip episodes on the $(\bar{1}01)$ forming $+60^{\circ}$ with the MRSS plane. 
More revealing is perhaps the case of glide when the MRSS is resolved on a $\{112\}$ plane --$(\bar{2}11)_{\rm T}$ to be precise--. In this case, Schmid behavior is generally recovered when eq.\ \ref{s} is used, as Fig.\ \ref{traj1} illustrates. The trajectory in this case follows a zig-zag pattern, characteristic of \emph{wavy} slip observed in bcc systems at low temperature ({\it e.g.}~\citet{franciosi1983}). However, non-Schmid behavior results in effective glide on the $(\bar{1}01)$ plane, forming $+30^{\circ}$ with the MRSS plane. This behavior is not inconsistent with recent Laue diffraction experiments of slip in W \citep{marichal2013} and with MD simulations performed with the same potential employed here by \citet{cere2013}.

At higher stresses and temperatures (cf.\ Fig.\ref{traj2}) the same general behavior can be observed, although the deviation from the MRSS plane for non-Schmid $\{110\}$ loading is more notable than under low stress/temperature conditions. In all cases, deviations from the MRSS plane are reliably in a counterclockwise direction. This is a direct manifestation of the twinning-antitwinning asymmetry that biases kink-pair nucleation toward planes that are consistent with the critical stresses shown in Fig.\ \ref{fig:schmid}. 

\subsection{Dislocation self-pinning}
\begin{figure}[h]
\centering
\fbox{\includegraphics[trim=1cm 4cm 3cm 5cm, clip=true, width=1.0\linewidth]{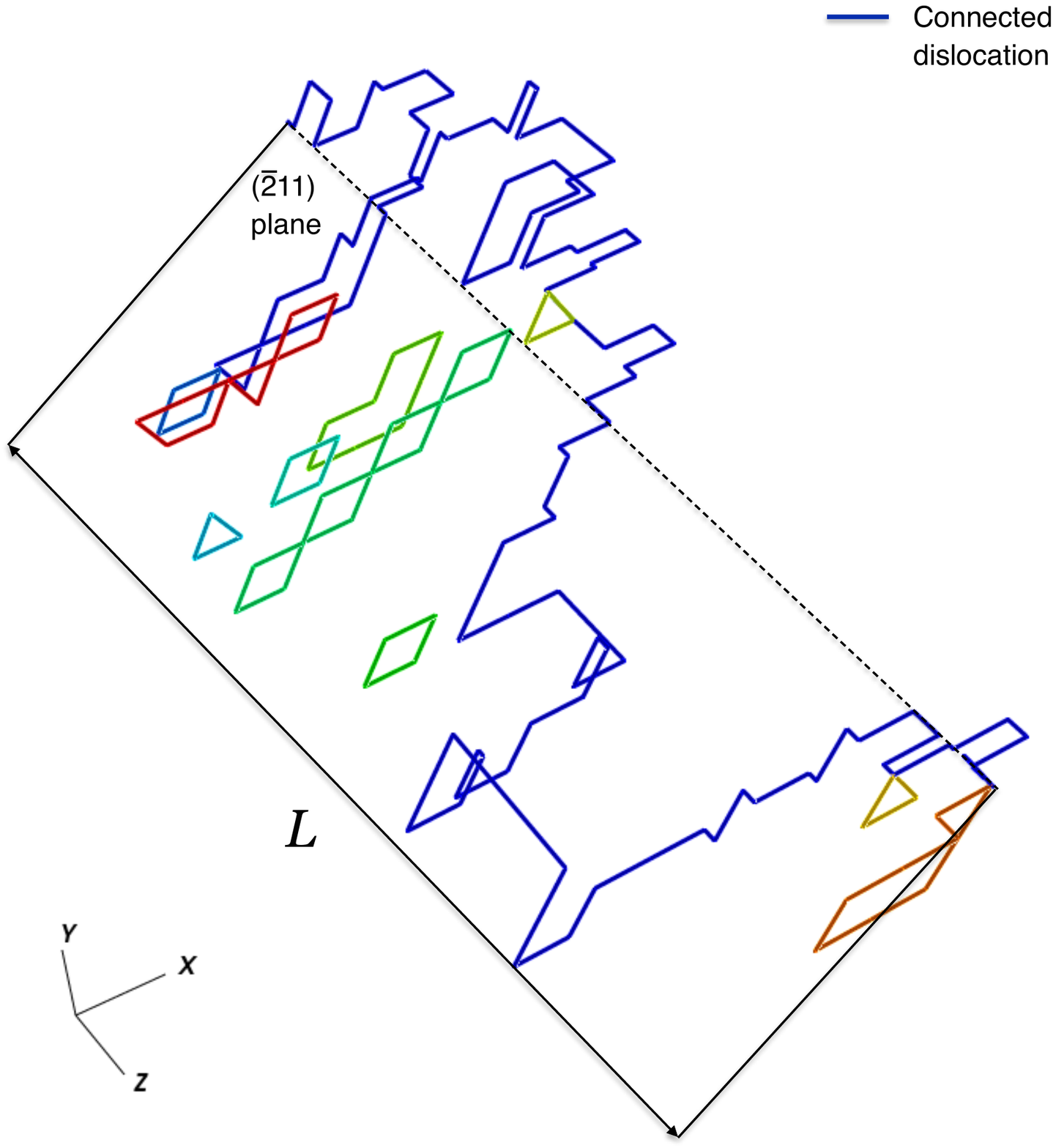}}
\caption{Final configuration after 5000 kMC cycles of a screw dislocation of length $L=1000b$ under $\{112\}$-Schmid loading conditions at 1000 MPa and a temperature of 1800 K.  Segments in dark blue belong to the main dislocation, while colored segments belong to detached loops. The depicted line configuration is scaled in the $z$ direction to facilitate viewing. See the Supplementary animation of the time evolution of the dislocation.\label{visit}}
\end{figure}

The reason for the loss of linearity in the $v$-$L$ dependence at high temperature or stress (Figs.\ \ref{subfig7} and \ref{subfig8}) is related to the increased probability of forming kink pairs on multiple glide planes simultaneously. According to equation \ref{req}, this probability increases with temperature, stress, and line length, consistent with the behavior discussed above. As alluded to in Section \ref{impl}, in multislip conditions the interaction among kink pairs on different planes results in cross kinks. These defects essentially halt the progress of the dislocation by acting as pinning points that must be overcome before motion can resume. When this happens, debris loops are formed in the wake of the main dislocation. 
Figure \ref{visit} shows the final configuration after 5000 kMC cycles of a screw dislocation of length $L=1000b$ under $\{112\}$-Schmid loading at 1000 MPa and a temperature of 1800 K. The figure clearly shows trailing chains of debris loops.
The reader is referred to the work by \citet{marian2004} for more details on the atomistic characteristics of this process. Here we quantify the formation of these loops and relate it to dislocation self-pinning and slowing down.

\begin{figure}[h]
\centering
  \includegraphics[width=1.0\linewidth]{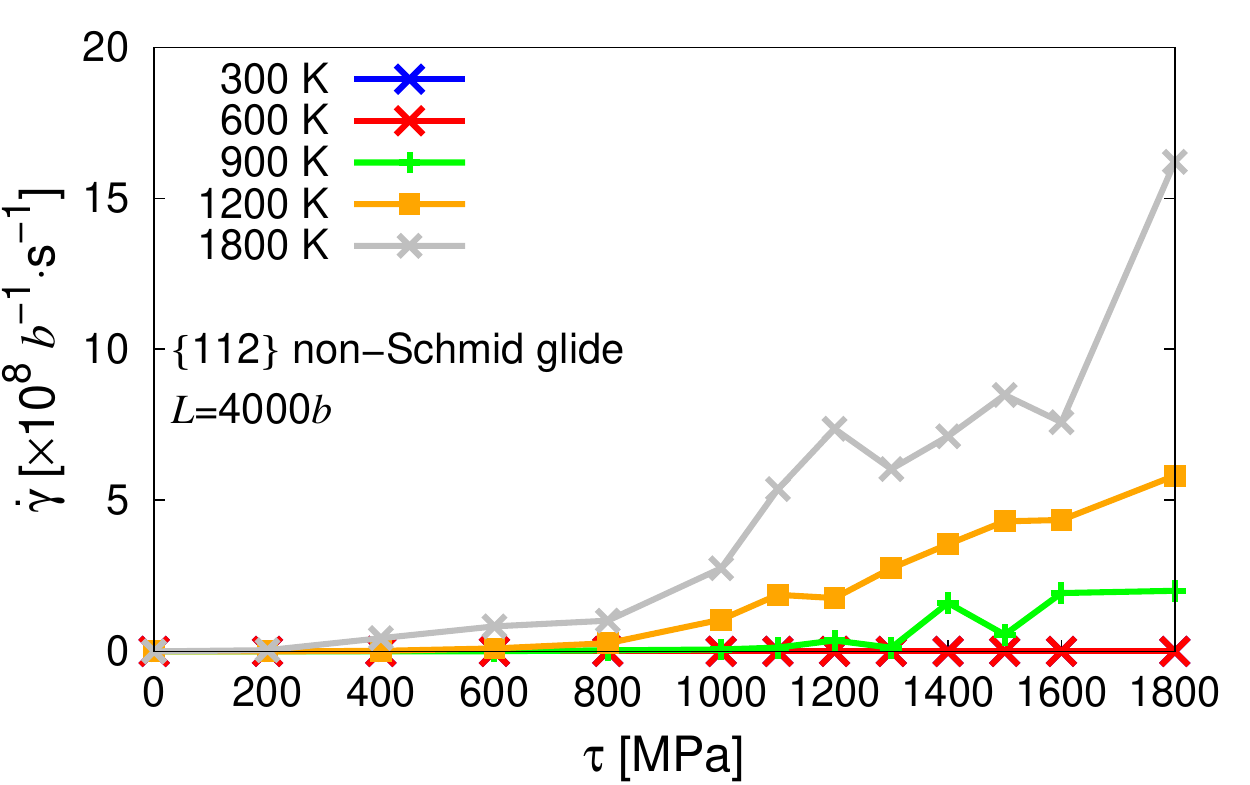}
   \caption{Debris loop generation rate as a function of temperature and applied stress for $\{112\}$ non-Schmid conditions for a dislocation with $L=4000b$. At 300 K there is zero loop generation.\label{debris1}}
\end{figure}

\begin{figure}[h]
\centering  
\includegraphics[width=1.0\linewidth]{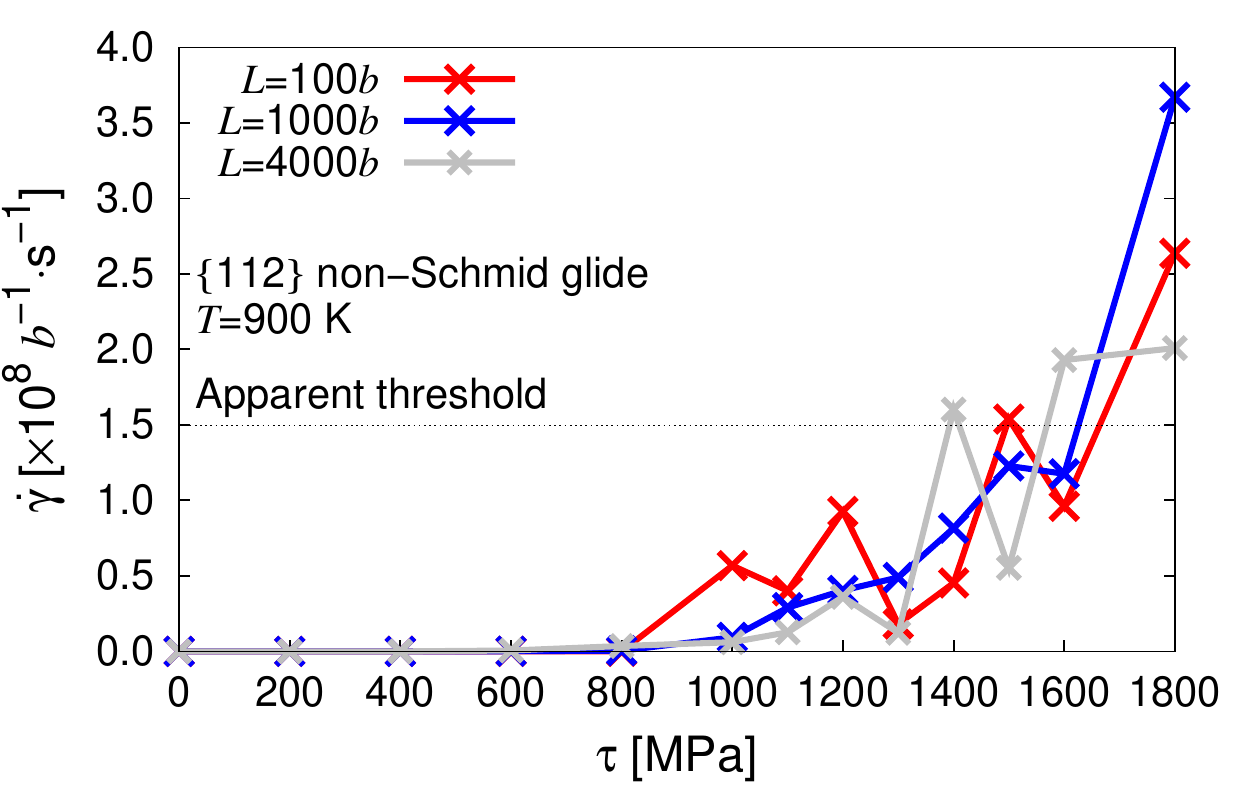}
\caption{Debris loop generation rate at 900 K during $\{112\}$ non-Schmid conditions for three different initial dislocation lengths. The apparent threshold above which self-pinning is seen to dominate the kinetics is marked with a dashed line.\label{debris2}}
\end{figure}

From analysis of trajectories such as that shown in Fig.\ \ref{visit}, the number of debris loops per unit time per unit length can be tallied as a function of $\tau$, $T$, $L$, and MRSS plane. This debris loop generation rate --which we term $\dot{\gamma}$-- is shown in Figure \ref{debris1} for $\{112\}$ non-Schmid loading for a dislocation with $L=4000b$. 
As expected, $\dot\gamma$ increases with increasing temperature and stress. However, for a given temperature and glide condition, the loop generation rate per unit length is independent of $L$. This is illustrated in Figure \ref{debris2}, where $\dot\gamma$ is shown as a function of stress for $\{112\}$ non-Schmid conditions and at 900 K for $L=100$, 1000, and $4000b$. In other words, the debris loop generation rate only depends on temperature, stress, and the glide condition. The example shown here is representative of other temperatures and MRSS plane orientations.

These results show that there may be a correlation between the degree of self-pinning in Figs.\ \ref{100b},  \ref{1000b}, and  \ref{4000b} and the value of $\dot\gamma$ for each case. For the specific example shown in Fig.\ \ref{debris2}, Fig.\ \ref{subfig6} suggests that the dislocation velocity deviates from the nominal exponential behavior at a stress of $\approx$1500 MPa. This corresponds to a value of $\dot\gamma=1.5\times10^8~{\rm s}^{-1}~b^{-1}$ (the approximate value of the three curves in Fig. \ref{debris2} at 1500 MPa). This apparent threshold is of course temperature dependent and varies with loading orientation, although here we only consider the case showcased in Figs.\ \ref{debris1} and \ref{debris2}. 

\subsection{Computational efficiency}
\begin{figure}[h]
\centering
\includegraphics[width=\linewidth]{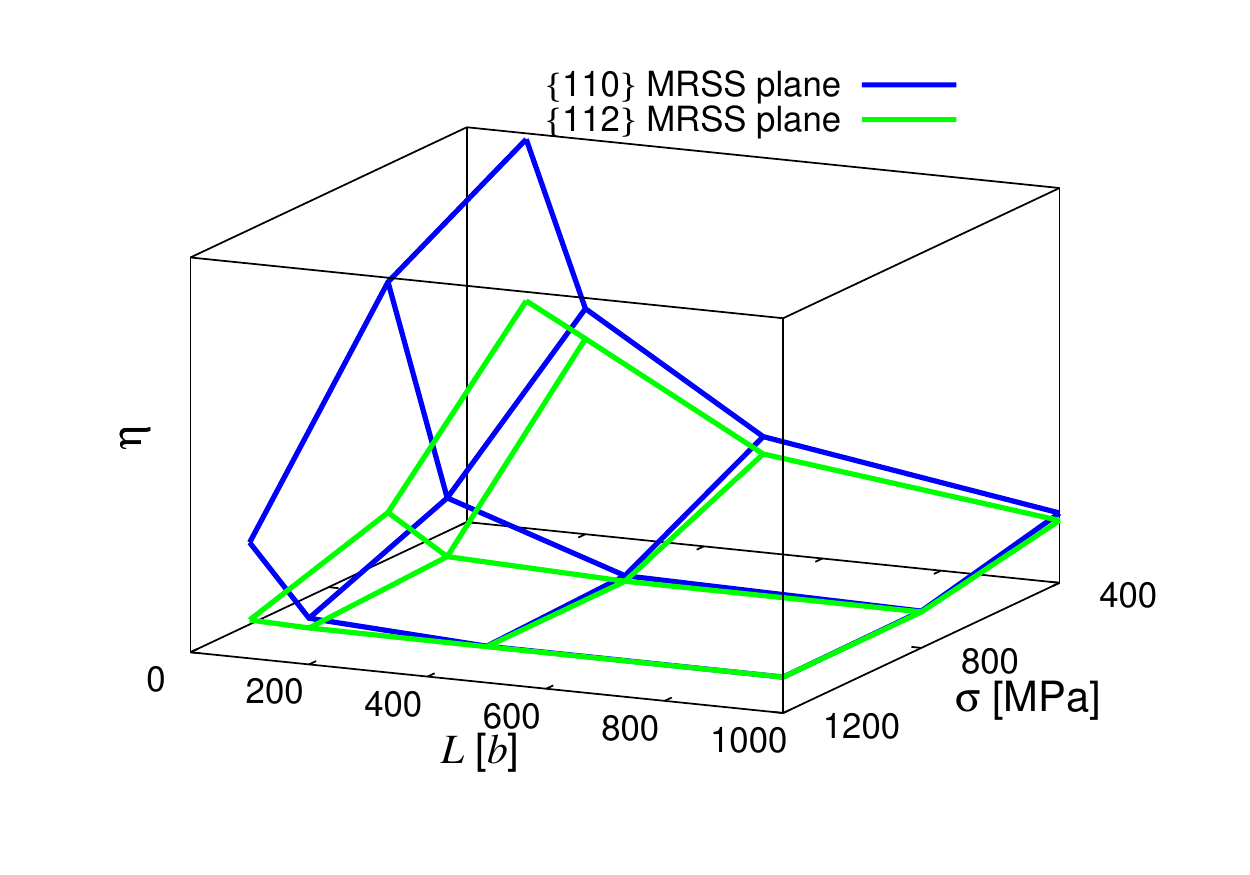}
\caption{Computational efficiency $\eta$ measured in arbitrary units as a function of $L$, $\tau$, and MRSS plane for a number of simulations conducted at 900 K for a fixed duration of 2000 kMC cycles.\label{perform}}
\end{figure}
As discussed in the previous section, the compounded effect of stress, temperature, and initial screw dislocation length, as well as stress orientation, is to enhance the probability to nucleate kink-pairs in multiple slip planes. This increases the number of segments and may lead to a stalled kinetic evolution as a consequence of self-pinning. Both of these phenomena decrease the computational efficiency of the kMC code, interpreted as the utilization of CPU time to result in net dislocation motion. The number of segments increases the numerical workload of the ${\cal O}(N^2)$ segment-segment stress calculation function\footnote{Profiling tests reveal that $>92\%$ of the CPU time in any given kMC cycle is spent in this function.}, while self-pinning arrests the dislocation progress resulting in slower net motion. To assess these overhead costs quantitatively, we plot in Figure \ref{perform} the dependence of the computational efficiency $\eta$ as a function of applied RSS and $L$. The calculations were performed for a fixed number of 2000 kMC cycles at 900 K. For clarity we display $\eta$ in arbitrary units to showcase the effect of each parameter studied, with quantitative details about the numerical values in each case given in \ref{a:eta}. As shown in the figure, increasing the stress, the dislocation line length, and/or under $\{112\}$ loading, all contribute to efficiency losses. Stress and temperature are generally equivalent in their effect on $\eta$, and so here only the impact of $\tau$ is evaluated.

\section{Discussion}
\emph{Motivation for using kMC simulations} -- The motion of screw dislocations proceeds via the thermally-activated nucleation of kink-pairs and kink propagation along the screw direction. Kinks are atomistic entities --as described in Figure \ref{isokink}-- but also elastic ones. This means that their properties must be characterized at the atomistic scale, but their effects can be potentially long-ranged. Dynamically, by virtue of its rare-event nature, kink pair nucleation operates on time scales that are hardly accessible by atomistic methods. This precludes, in most cases the use of direct MD or other atomistic methods.
However, dislocations containing kink pairs are subjected to long-range elastic self-forces, which have to be integrated along the dislocation in order to be evaluated and resolved spatially. As this is typically very numerically-intensive, we resort to discretization methods that treat dislocation lines as piece-wise entities in which all segments interact with all segments. This, for its part, precludes the use of effective-medium methodologies such as the line-tension approximation or other techniques in which these $O(N^2)$ interactions are not captured.
KMC, in our mind, offers the ideal alternative to bridge these two limits. On the one hand, the dislocation is treated as a piece-wise object attached to an underlying lattice. This allows us to represent some of the most important atomistic features of the dislocation fairly accurately. At the same time, this piece-wise representation enables the calculation of all the elastic forces in an efficient manner. The result is a method that can access time scales long enough to statistically capture dislocation motion, yet it retains sufficient detail to accurately provide a clear connection to the underlying atomistic physical features.

\begin{figure}[h]
\centering
\includegraphics[width=1.0\linewidth]{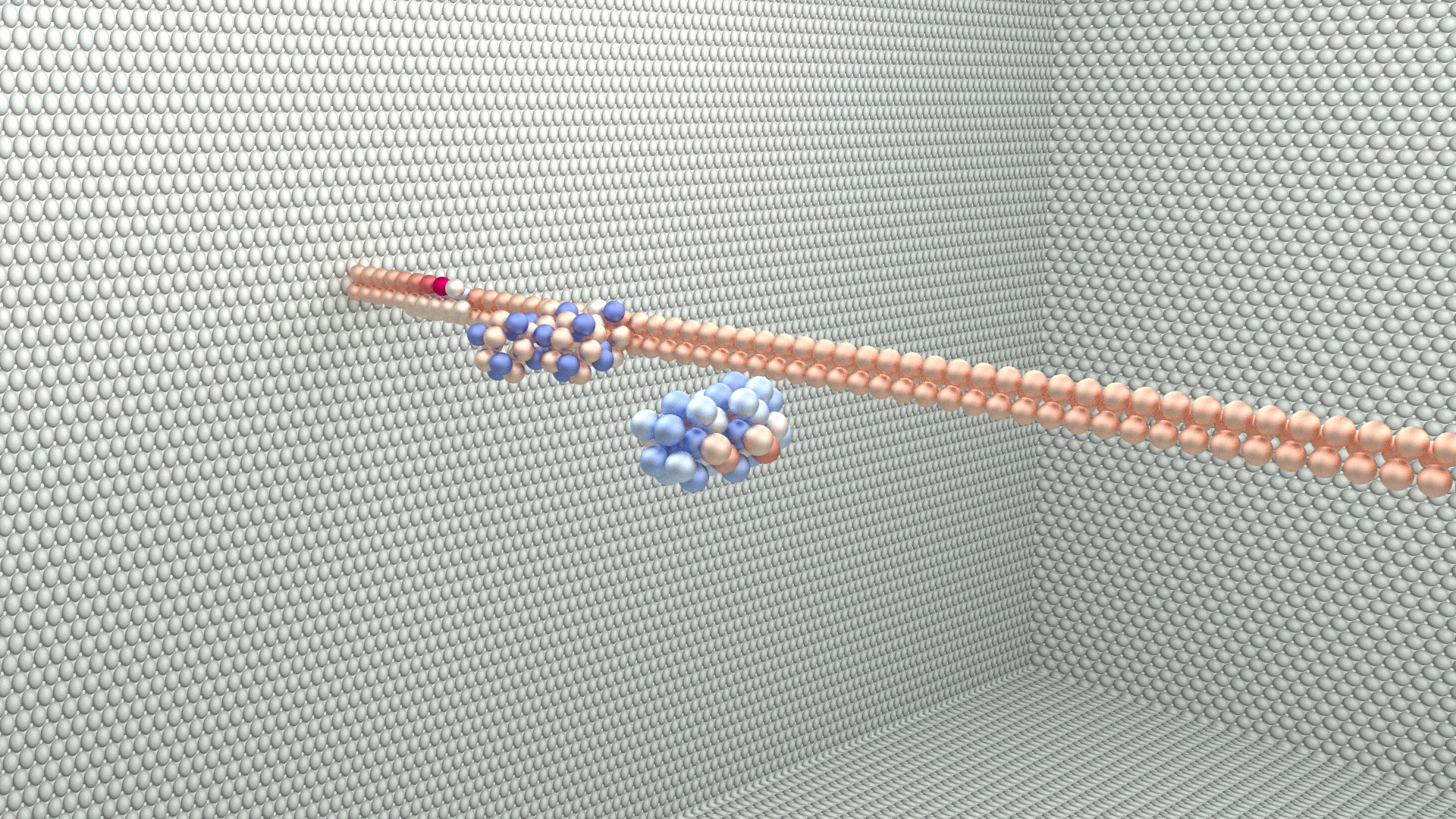}
\caption{MD simulation of a screw dislocation under the following conditions: $L=100b$, $T=300$ K, $\sigma_{\rm RSS}=1100$ MPa, MRSS plane $\equiv\{112\}$. After a few time steps, the dislocation starts producing debris in the form of vacancy and interstitial clusters. These are akin to small dislocation loops in he kMC simulations.\label{MD}}
\end{figure}

\emph{Comparison with MD results} -- One of the main motivations behind the development of our kMC model was MD's inability to sample thermally activated motion within its space and time limitations. It is then useful to compare MD and kMC results of screw dislocation glide subjected to nominally identical boundary conditions. However, as discussed above, the overdriven nature of MD simulations causes the occurrence of cross-kinks and associated debris for line lengths for which the kMC simulations predict smooth glide. This is illustrated in Figure \ref{MD}, where a screw dislocation of length $100b$ is seen to leave vacancy clusters behind at 300 K and 1100 MPa of stress applied on a $\{112\}$ plane. For the current interatomic potential, the threshold length below which cross kinks are not seen to occur was estimated to be $25b$ \citep{cere2013}. This is below the length for which kMC simulations can support an elementary kink pair. Therefore, we are forced to make an imperfect comparison between the MD results with $L=25b$ and the kMC results for $L=75b$, which is near the minimum length in kMC calculations to contain one kink-pair. 

\begin{figure}[h]
\centering
\includegraphics[width=1.0\linewidth]{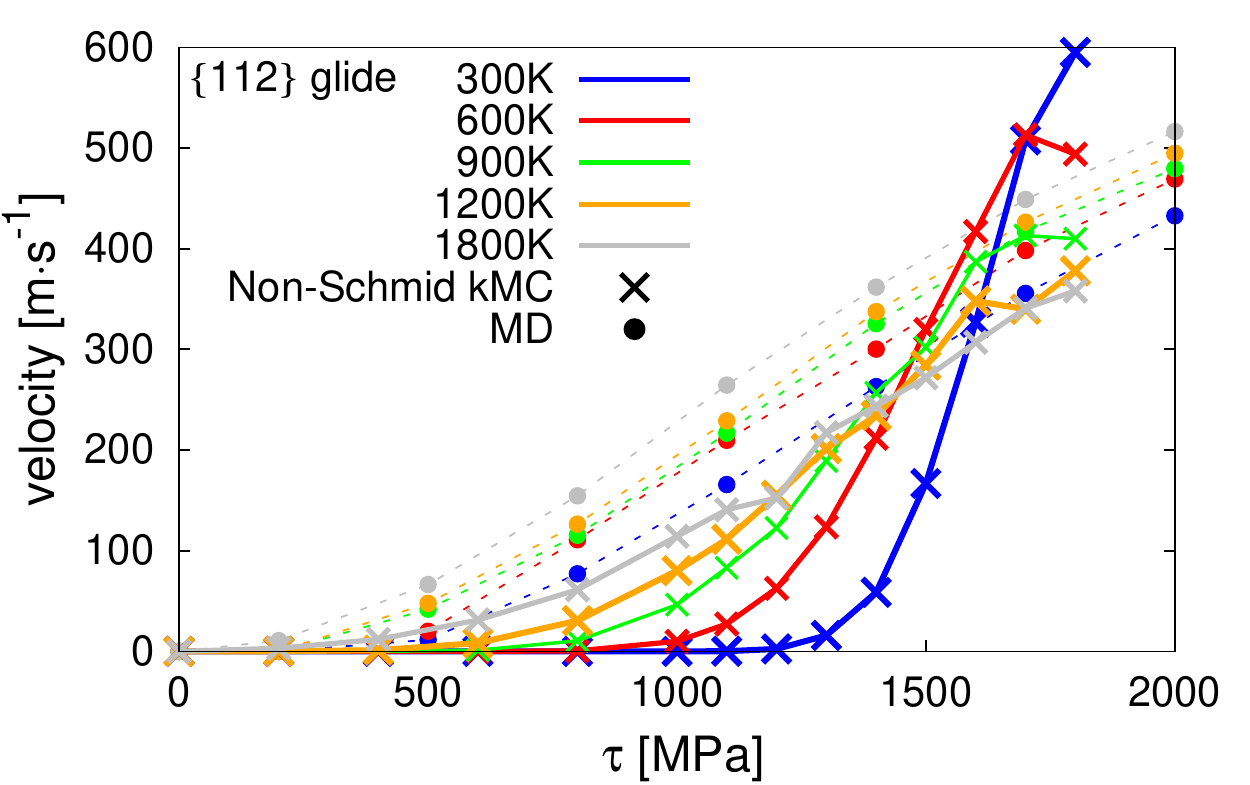}
\caption{Comparison of dislocation velocities from MD results \citep{cere2013} and kMC calculations.\label{MD2}}
\end{figure}

Results from both approaches are shown in Figure \ref{MD2}. The figure shows that the MD velocities are systematically higher than their kMC counterparts below 1500 MPa. Above this value, the kMC velocities at 300 and 600 K overtake the MD-calculated values. It is interesting to note that the qualitative shape of the MD curves coincides with those of the kMC curves at the highest temperatures of 1200 and 1800 K. This is symptomatic of the limitations of MD, which even at low stresses and temperatures create simulation conditions that are representative of higher values. It must also be kept in mind that a sensitivity study has not been conducted on the kMC parameters, and thus the present comparison is only valid inasmuch as the current parameterization can be considered a sufficiently valid one for the method. In terms of computational overhead, MD simulations are approximately three to seven orders of magnitude costlier than their kMC counterparts on the basis of the metric employed in Table \ref{tbs}. We refer the reader to \ref{a:eta} for more details.

\emph{Dislocation self-pinning} -- Self-pinning occurs as a consequence of the formation of cross-kinks, which act as strong sessile junctions. Cross-kinks may be resolved topologically by complementary kink pairs, resulting in the closing of a \emph{debris} loop behind. Loop generation contributes to self-pinning as well. 
The energy expended in producing debris loops is taken out of the total mechanical work available to make the dislocation glide, which results in an effective `reduced' stress and, therefore, lower velocities. Physically, self-pinning is seen to become important above a certain generation rate threshold, which correlates with a leveling-off of dislocation velocity curves as a function of stress.

This notion of \emph{threshold} generation rate originates in the creation of kink-pairs on multiple slip planes, whose effect in the kinetic behavior depends on the combined effects of cross-kink production and resolution. An enhanced probability of kink pair production (brought about by increasing temperature, stress, and/or multislip conditions) may facilitate the production of cross-kinks, leading to potentially higher self-pinning. At the same time, the probability for resolution of these is also intensified by the same processes. Resolution of cross kinks results in debris loop production. Beyond the apparent debris generation threshold, however, the production of cross-kinks overruns the likelihood of resolution, effectively arresting the dislocation progress and stagnating the velocity increase with temperature and stress. When this happens, debris production is simply a manifestation of self-pinning on the larger scale.
This is one of the reasons leading to the length independent behavior observed at mid-to-high temperatures and stresses (cf.~Fig.\ \ref{length2}), and which is behind the anomalous behavior of some dislocations observed experimentally \citep{luke2007}.

\begin{figure}[h]
\centering
\includegraphics[width=1.0\linewidth]{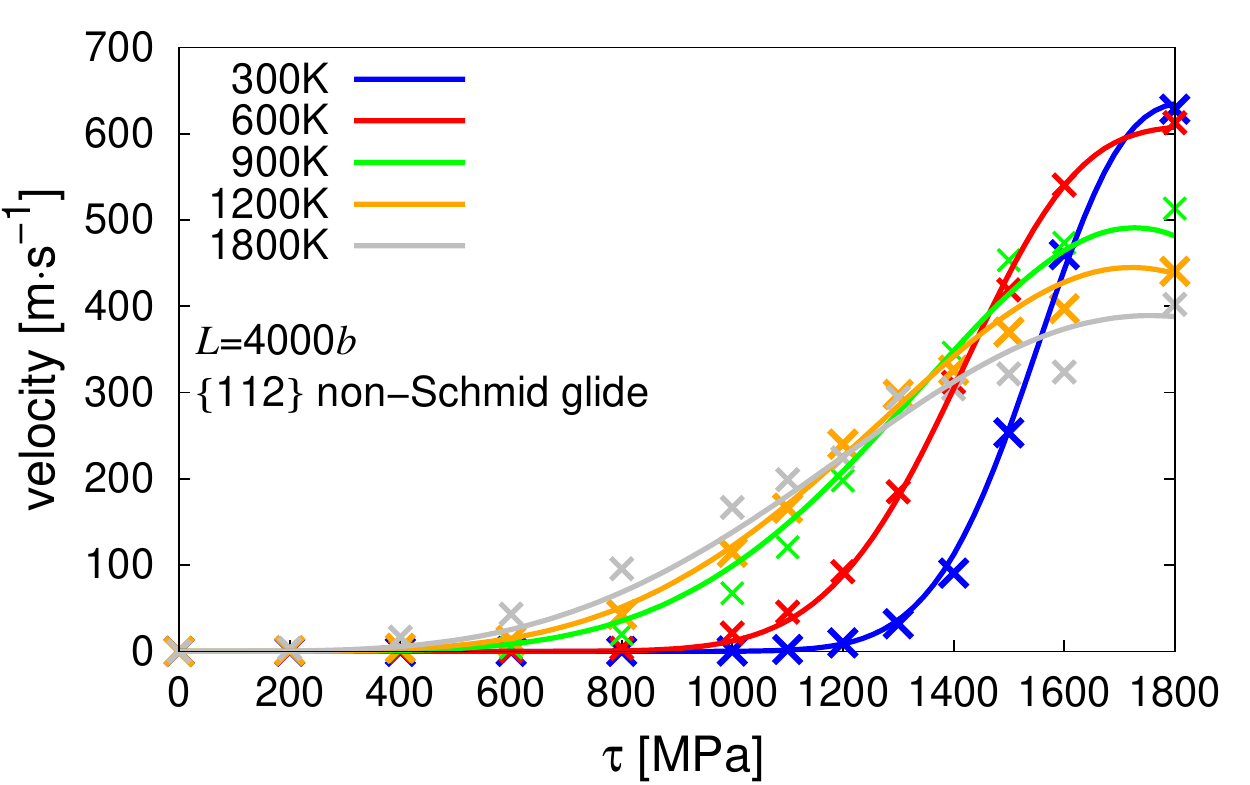}
\caption{Comparison between eq.\ \ref{fitt} (solid lines) parameterized for $L=4000b$ under non-Schmid $\{112\}$ glide conditions and the actual kMC data.\label{figfit}}
\end{figure}

\emph{Extraction of effective mobility laws} -- Ultimately, the data compiled in this work via extensive kMC calculations must be used to fit mobility functions suitable for, {\it e.g.} dislocation dynamics, phase field, or crystal plasticity simulations (see for example \citet{tang2014}). The deviations exposed by our calculations from the expected exponential behavior due to self-pinning call for a possible fitting function of the following type:
\begin{equation}
v(s,T) = A's^{n'} f'(s, T)\left(1 - B' f'(s, T) \right )\\
\label{fitt}
\end{equation}
$$f'(s, T) = \exp\left\{ -\frac{\Delta H_0}{kT} \left( 1 - s^{p'}  \right )^{q'} \right\}$$
where $A'$, $B'$, $n'$, $p'$, and $q'$ are all adjustable parameters and $s$ is defined as in eq.\ \ref{s} or \ref{a4}. The above expression captures the leveling-off displayed in the $v$-$\tau$ relations at high stress and temperature.
By way of example, here we fit the results for $L=4000b$. Table \ref{ttt} gives the parameters under each specific glide condition.
\begin{table}[h]
\caption{Adjustable parameters for the fitting function given in eq.\ \ref{fitt}. The units of $A'$ are such that $v(s,T)$ is in m$\cdot$s$^{-1}$, {\it i.e.}~m$\cdot$s$^{-1}$$\cdot$MPa$^{-n}$. All other parameters are non-dimensional.}
\label{ttt}
\centering
\begin{tabular}{|c|ccccc|}
\hline
Temperature range [K] & $A'$ &  $n'$ &  $B'$  & $p'$ & $q'$ \\
\hline
\multicolumn{6}{|c|}{$\{110\}$ Schmid loading}\\
\hline
All temperatures  & 3693.4 & 2.47 & 0.97 & 0.16 & 1.00 \\
\hline
\multicolumn{6}{|c|}{$\{110\}$ Non-Schmid loading}\\
\hline
$300$  & 698.2 & 0.30 & 0.0 & 1.15 & 2.97 \\
$>300$  & 1444.2 & 1.78 & 0.72 & 0.26 & 1.40 \\
\hline
\multicolumn{6}{|c|}{$\{112\}$ Schmid loading}\\
\hline
All temperatures  & 755.6 & 0.38 & 0.50 & 0.22 & 1.01 \\
\hline
\multicolumn{6}{|c|}{$\{112\}$ Non-Schmid loading}\\
\hline
$\leq600$  & 2084.2 & 1.39 & 0.68 & 0.81 & 2.45 \\
$>600$  & 3416 & 2.72 & 0.89 & 0.19 & 1.32 \\
\hline
\end{tabular}
\end{table}
Figure \ref{figfit} shows the fit for non-Schmid conditions on a $\{112\}$ plane. The agreement between the fitting functions and the data is similar for other glide conditions and/or values of $L$.

\emph{Limitations of the method} -- We conclude this section discussing some of the limitations of our model. First, the sampling function \ref{req} contains several parameters with exponential dependence that have been obtained via atomistic calculations using a recent interatomic potential. As such, they are subjected to errors associated with the atomistic technique used (NEB), the type of potential and its parameterization (EAM), and the least-squares fitting procedure. In a way, all these errors are unavoidable --in the sense that we have employed `state-of-the-art' techniques and procedures-- but their impact on the overall kinetics, although unassessed at the moment, might conceivably be notable in some cases.
Next, the very physical foundation of the code --the Arrhenius expression for the thermally activated kink-pair nucleation rate-- may be called into question under some of the conditions explored here. Indeed, at high stresses (and temperatures) the kinetics is better represented by generalized Arrhenius forms, {\it e.g.} the Jackson formula \citep{swinburne2013b}, and this may affect the high stress/temperature tails of the velocity-stress relations given in Figs.\ \ref{100b}, \ref{1000b}, and \ref{4000b}.
The representation of dislocation segments may also be a source of errors in our setting. Kinks and screw segments are joined by sharp corners that give rise to stress singularities --these are avoided here by resorting to a \emph{screening} distance within which the stress is not calculated-- that are artifacts of our piecewise rectilinear representation of dislocation lines.
Another physical phenomenon not captured in these simulations is the softening of the elastic constants and Peierls (critical) stress with temperature. In particular, today's computational resources permit the direct calculation of the temperature dependence of the critical stress \citep{gilbert2013}. It is not clear at his point how significant this dependence is on the dislocation velocities calculated here.
Finally, it is worth mentioning that the impact on dislocation motion of non-glide stresses --another source of non-Schmid effects-- is not presently considered in this work, although its implementation is straightforward if the data were available.

\section{Summary}
We have developed a kinetic Monte Carlo model of thermally-activated screw dislocation motion in bcc crystals, with a current parameterization for W using a state-of-the-art interatomic potential. Our method includes all relevant physical processes attendant to screw dislocation motion, including --for the first time-- kink diffusion and non-Schmid effects. 

With the versatility and efficiency afforded by our kMC algorithm, we have studied dislocation mobility as a function of stress, temperature, initial dislocation line length, and MRSS plane orientation. An attractive feature of the present calculations is that they allow us to separate important mobility dependencies and assess their impact on the kinetics individually.

We find that non-Schmid effects have an important influence on the absolute value of the velocity as function of both stress and temperature, suggesting that they cannot be neglected in plasticity simulations. We also find that at sufficiently high stresses and temperatures, self-pinning processes control dislocation motion. Finally, some effective fitting functions are proposed that capture the essential features of dislocation motion to be used in more homogenized models of crystal deformation.

\section*{Acknowledgments}
We are indebted to V. Bulatov for useful discussions and helpful guidance. Conversations with D. Rodney, M. Gilbert, and W. Cai are gratefully acknowledged.
This work was performed under the auspices of the U.S. Department of Energy by Lawrence Livermore National Laboratory under Contract No.\ DE-AC52-07NA27344. J. M. acknowledges support from DOE's Early Career Research Program. 
D. C. acknowledges support from the Consejo Social and the PhD program of the Universidad Polit\'ecnica de Madrid.
T. D. S. was supported through a studentship in the Centre for Doctoral Training on Theory and Simulation of Materials at Imperial College London funded by EPSRC under Grant No. EP/G036888/1.
This work was part-funded by the RCUK Energy Programme (Grant Number EP/I501045) and by the European Union's Horizon 2020 research and innovation programme. The views and opinions expressed herein do not necessarily reflect those of the European Commission. This work was also part-funded by the United Kingdom Engineering and Physical Sciences Research Council via a programme grant EP/G050031.

\clearpage
\appendix
\section{Computing diffusion and drift coefficients of isolated single kinks}
\label{a:kink}
\begin{figure}[h]
\includegraphics[width=0.8\linewidth]{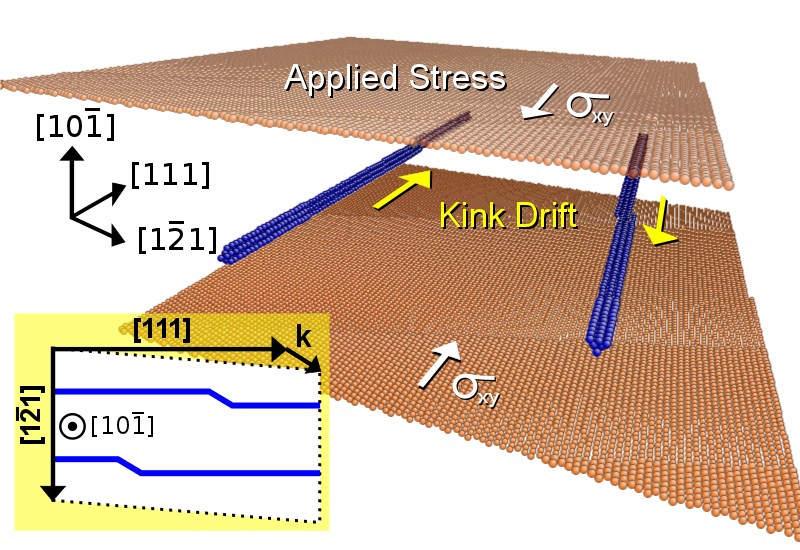}
\caption{Illustration of kink drift simulations. Kinks on a $\small{\nicefrac{1}{2}}\langle111\rangle\{10\bar{1}\}$ screw dislocation dipole, characterized by a lattice `kink' vector $\vec{k}$, are subject to an applied stress on bounding $(10\bar{1})$ planes. Under no applied stress with fully periodic boundary conditions the kinks diffuse freely. Inset: Cartoon of the supercell along $[10\bar{1}]$, illustrating the relation of the kink vector to a kinked dislocation line.\label{setup}}
\end{figure}

To generate isolated kinks in an MD supercell, we use especial boundary conditions that enforce a tilt equal to a lattice vector $\vec{k}$. Depending on the value of $\vec{k}$ kinks of opposite signs --`right' and `left', to employ the usual convention-- are created in cells containing a balanced dislocation dipole.
These configurations are then equilibrated at finite temperature and the simulation output is then time averaged and energy filtered in both zero and finite stress conditions to produce a series of kink positions $x$ from which a kink drift and diffusivity can be statistically determined. This procedure is described in detail by Swinburne \etal~\citep{swinburne2013}, and a typical simulation supercell (containing around $10^6$ atoms) is depicted in Figure \ref{setup}.

\begin{figure}[h]
\includegraphics[width=\linewidth]{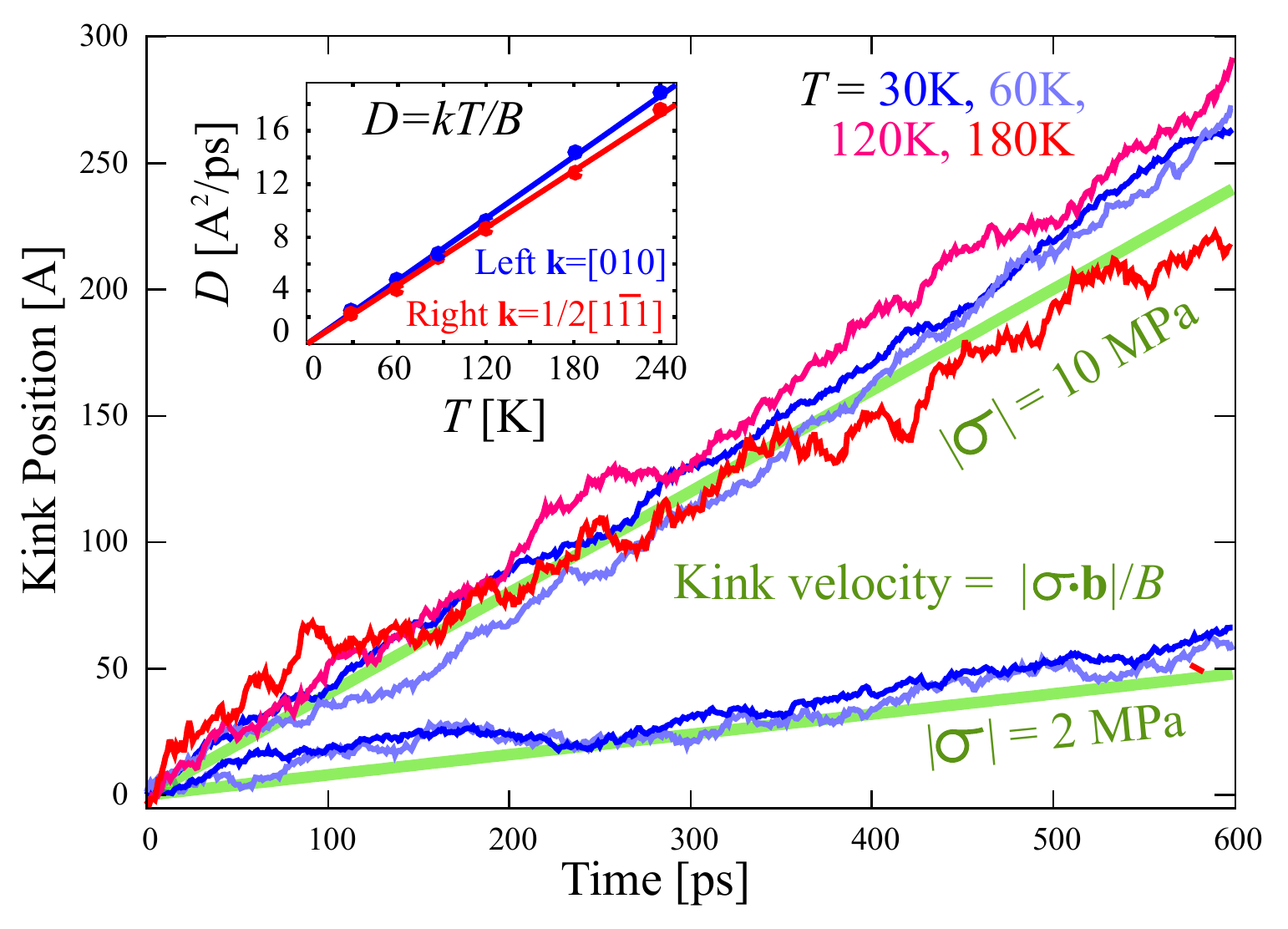}
\caption{Results of kink drift simulations for $\vec{k}=\small{\nicefrac{1}{2}}[1\bar{1}1]$ (right) kinks on $\small{\nicefrac{1}{2}}\langle111\rangle\{10\bar{1}\}$ screw dislocations. We see a temperature independent drift velocity $v_k=|\matr{\sigma}\cdot\vec{b}|/B$ in very good agreement with $B$ determined from zero stress kink diffusion simulations (green lines). Inset: Results from kink diffusion simulations. We see the diffusivity $D=kT/B$ rises linearly with temperature, meaning that $B$ is independent of temperature.\label{kd_res}}
\end{figure}

The results of these simulations are displayed in Figure \ref{kd_res}. Kinks were observed to freely diffuse with a diffusivity $D=kT/B$ under no applied stress with fully periodic boundary conditions, while, under stresses of 2$\sim$10 MPa applied to the bounding $(10\bar{1})$ planes, kinks were observed to drift with a viscous drag law $\dot{x}=|\matr{\sigma}\cdot\vec{b}|/B$. Although the two screw dislocations eventually annihilate under applied stress, for a sufficiently wide and long supercell, the kinks drift independently for at least two supercell lengths ($\sim$600 \AA) before any influence of their mutual attraction can be detected.

The drift and diffusion simulations are fitted to the Einstein relation:
$$D=kT\lim_{|\matr{\sigma}\cdot\vec{b}|\to0}\frac{\dot{x}}{|\matr{\sigma}\cdot\vec{b}|}$$ 
whereupon it is seen that the viscous drag $B$ is independent of temperature and shows little variation between left and right kinks. The final mobility laws were determined to be $v_k=3.8\times10^{-6}\tau$ for $\vec{k}=\small{\nicefrac{1}{2}}[1\bar{1}1]$ (`right' or `interstitial') kinks and $v_k=4.0\times10^{-6}\tau$ for $\vec{k}=[010]$ (`left' or `vacancy') kinks. These velocities are in m$\cdot$s$^{-1}$ when the stress is in Pa. Phonon scattering treatments \citep{hirth} predict that $B$ should increase linearly with temperature due to the increased phonon population, but the observed temperature independence of $B$ agrees with previous studies of kink diffusion \citep{swinburne2013} and other nanoscale defects \citep{Dudarev2008}. The assumption of constant kink velocity is justified on the basis of the energy landscape over which kinks move. This landscape is effectively flat --albeit with significant thermal roughness-- due to the onset of vibrational chaos, which destroys inertial motion and leaves linear viscous motion as the dominant one. Potentially, at very low temperatures and high stresses this regime could break down and be replaced by one where inertial effects are dominant. However, we have not contemplated this possibility. 

\section{Implementing non-Schmid effects in the kinetic Monte Carlo calculations}
\label{a:schmid}

\begin{figure}[h]
\centering
\includegraphics[width=1.0\linewidth]{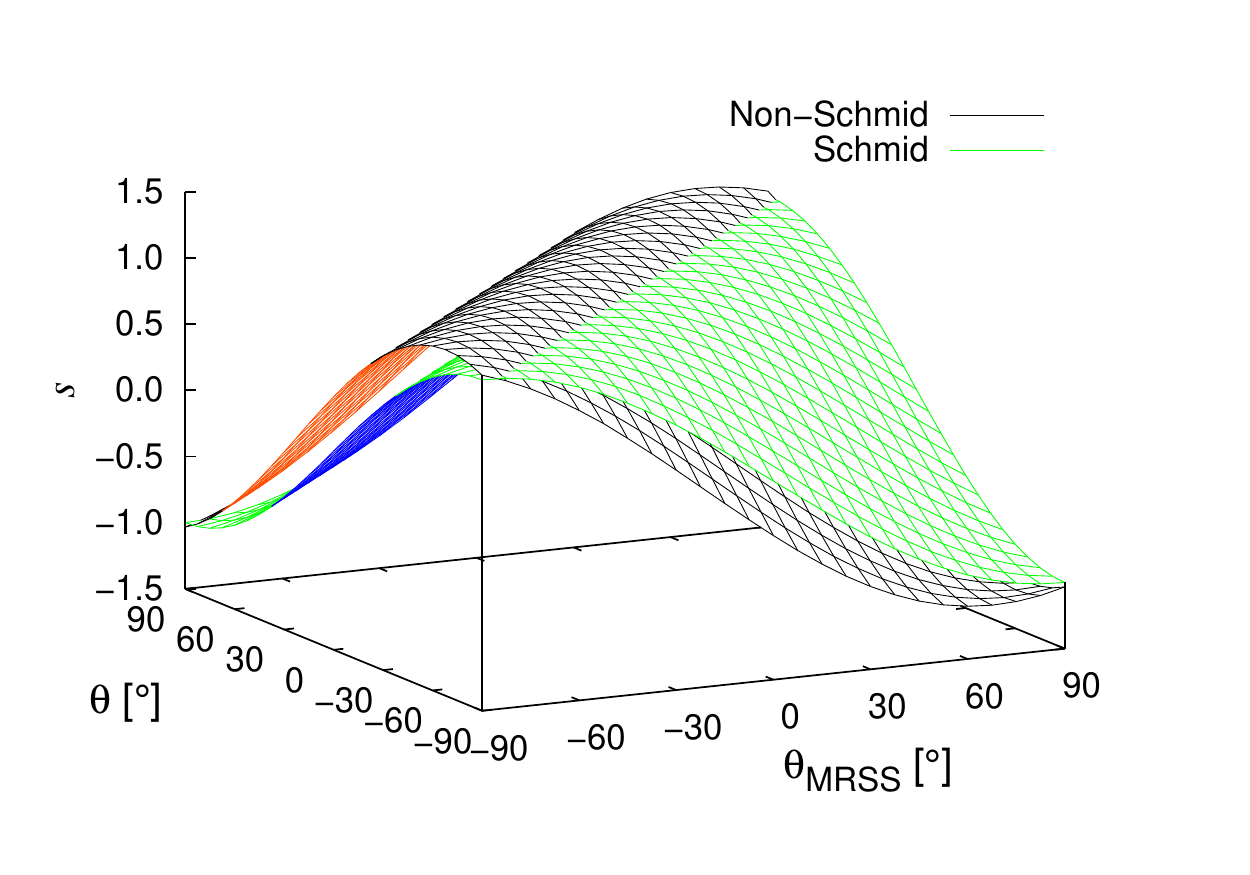}
\caption{Comparison between the normalized stress $s$ under Schmid and non-Schmid conditions as a function of $\theta$ and $\theta_{\rm MRSS}$. Recall that $\chi=\theta_{\rm MRSS}-\theta$.\label{schhh}}
\end{figure}

In the reference system used in Fig.\ \ref{110}, the MRSS is unequivocally defined as:
$$\sigma_{\rm MRSS}=\sqrt{\sigma_{xz}^2+\sigma_{yz}^2}$$
with
$$\theta_{\rm MRSS}=\arctan{\left(-\frac{\sigma_{xz}}{\sigma_{yz}}\right)}$$
and 
$$\chi=\theta_{\rm MRSS}-\theta$$
For the purpose of the implementation of non-Schmid effects, we express eq.\ \ref{s} in terms of the MRSS by noting that, from Fig.\ \ref{110}, $\sigma_{\rm RSS}=\sigma_{\rm MRSS}\cos\chi$:
\begin{equation}
s(\chi)=\frac{\sigma_{\rm MRSS}\cos\chi}{\sigma_P}
\label{a1}
\end{equation}
Schmid law states that the critical stress $\sigma_c(\chi)$ depends on $\chi$ as:
\begin{equation}
\sigma_c(\chi)=\frac{\sigma_P}{\cos\chi}
\end{equation}
which results in rewriting eq.\ \ref{s} as:
\begin{equation}
s(\chi)=\frac{\sigma_{\rm MRSS}}{\sigma_c(\chi)}
\label{a3}
\end{equation}
Proving that eqs.\ \ref{s} and \ref{a3} are equivalent is straightforward:
$$
\begin{aligned}
\sigma_{\rm RSS} &= \sigma_{\rm MRSS}\cos\chi \\
				&= \sigma_{\rm MRSS}\cos(\theta_{\rm MRSS}-\theta) \\
				&= \sigma_{\rm MRSS}\left[\sin\theta_{\rm MRSS}\sin\theta+\cos\theta_{\rm MRSS}\cos\theta\right] \\
				&= \sigma_{\rm MRSS}\sin\theta_{\rm MRSS}\sin\theta+\sigma_{\rm MRSS}\cos\theta_{\rm MRSS}\cos\theta \\
				&= -\sigma_{xz}\sin\theta+\sigma_{yz}\cos\theta
\end{aligned}
$$
From this, non-Schmid effects are introduced by substituting the following expression:
$$\sigma_c(\chi)=\frac{a_1\sigma_P}{\cos\chi+a_2\cos\left(\pi/3+\chi\right)}$$
into eq. \ref{a3}:
\begin{equation}
s(\chi)=\frac{\sigma_{\rm MRSS}}{\sigma_c(\chi)}=\frac{\sigma_{\rm MRSS}\left(\cos\chi+a_2\cos\left(\pi/3+\chi\right)\right)}{a_1\sigma_P}
\label{a4}
\end{equation}
whence it is readily seen that Schmid behavior is recovered for $a_1=1$ and $a_2=0$. Figure \ref{schhh} showcases the difference between $s(\chi)$ for Schmid and non-Schmid behavior as a function of $\theta$.

\section{Computational efficiency}
\label{a:eta}
The computational efficiency is assessed in the following manner. For the purposes of this paper, we assume that the productivity of a kMC run is based on the distance traveled by a dislocation during a fixed number of cycles, as a longer distance results in better converged velocity calculations and more precise data. Our performance metric of choice is then to normalize the distance traveled in each case by the CPU time invested in achieving it. Table \ref{tbs} gives the numerical values for this metric in \AA~per second of CPU time for various dislocation lengths and applied stresses. These data are the basis for what is shown in Fig.\ \ref{perform}.

\begin{table}[h]
\caption{Numerical values in \AA~per CPU second of the efficiency metric considered to evaluate the kMC code's performance under different conditions.}
\label{tbs}
\centering
\begin{tabular}{ccccc}
\hline
\multicolumn{5}{c}{$\tau=400$ MPa, $T=900$ K, 2000 kMC steps}\\
\hline
$L$ &  $100b$ &  $200b$  & $500b$ & $1000b$ \\
\hline
$\{110\}$ MRSS plane & 1180  &  634 & 267 & 113 \\
$\{112\}$ MRSS plane  & 640 & 534 & 209 & 88 \\
\hline
\end{tabular}
\begin{tabular}{ccccc}
\hline
\multicolumn{5}{c}{$\tau=800$ MPa, $T=900$ K, 2000 kMC steps}\\
\hline
$L$ &  $100b$ &  $200b$  & $500b$ & $1000b$ \\
\hline
$\{110\}$ MRSS plane & 921  &  219 & 20.0 & 2.7 \\
$\{112\}$ MRSS plane  & 150 & 23.0 & 1.0 & - \\
\hline
\end{tabular}
\begin{tabular}{ccccc}
\hline
\multicolumn{5}{c}{$\tau=1200$ MPa, $T=900$ K, 2000 kMC steps}\\
\hline
$L$ &  $100b$ &  $200b$  & $500b$ & $1000b$ \\
\hline
$\{110\}$ MRSS plane & 267  &  34.7 & 2.0 & 0.21 \\
$\{112\}$ MRSS plane  & 8.3 & 1.7 & - & - \\
\hline
\end{tabular}

\end{table}

As a point of comparison with `equivalent'\footnote{In the sense that they are designed to measure similar properties.} MD simulations, we first resort to the data published by \citet{cere2012}, where the nominal cost of one time step per atom is $\approx1.5\times10^{-5}$ CPU seconds for the interatomic potential employed here. For 750,000 atoms, that is 11.25 CPU s per time step. Typical MD simulations involve $10^5$ steps of 1 fs each, which results in $1.12\times10^6$ CPU seconds. Again, per the data in ref.\ \citep{cere2012}, those simulations achieve displacements on the order of 850~\AA, which results in $7.5\times10^{-4}$ \AA~per CPU second. This represents efficiencies of three to seven orders of magnitude lower than our kMC simulations.



\bibliographystyle{model5-names}





\end{document}